\documentclass[
showpacs, preprintnumbers,
nofootinbib, aps, prd, superscriptaddress,
10pt, showkeys, notitlepage, longbibliography]{revtex4-1}

\usepackage[dvips]{graphicx}
\usepackage{bm,latexsym,amsmath,amssymb,amsfonts,times}
\usepackage{color}
\usepackage{natbib}
\usepackage{times}
\usepackage{graphicx}
\usepackage{dcolumn}
\usepackage{mathtools}

\def\be {\begin{equation}}
\def\ee {\end{equation}}
\def\ba {\begin{eqnarray}}
\def\ea {\end{eqnarray}}

\def\bi {\begin{itemize}}
\def\ei {\end{itemize}}

\newcommand{\cR}{{\cal R}}

\newcommand\beq{\begin{eqnarray}}
\newcommand\eeq{\end{eqnarray}}

\newcommand\bA{{\bar A}}
\newcommand\bF{{\bar F}}

\usepackage[linktocpage,breaklinks]{hyperref}
\definecolor{venetianred}{rgb}{0.78, 0.03, 0.08}
\definecolor{grey}{rgb}{0.25, 0.25, 0.28}
\definecolor{darkmidnightblue}{rgb}{0.0, 0.2, 0.4}
\definecolor{egyptianblue}{rgb}{0.06, 0.2, 0.65}
\definecolor{darkblue}{rgb}{0.0, 0.0, 0.55}
\hypersetup{colorlinks,breaklinks,
            linkcolor=red,urlcolor=darkblue,
            anchorcolor=red,citecolor=blue}

\def\X5sp{{\rm X}_5}
\def\Y3sp{{\rm Y}_3}
\def\Z3sp{{\rm Z}_3}

\begin{document}

\title{Proca stars with nonminimal coupling to the Einstein tensor}

\author{Masato Minamitsuji}
\email{masato.minamitsuji@ist.utl.pt}
\affiliation{Centro Multidisciplinar de Astrofisica - CENTRA,
Instituto Superior Tecnico - IST,
Universidade de Lisboa - UL,
Avenida Rovisco Pais 1, 1049-001, Portugal.}

\begin{abstract}
We investigate Proca star (PS) solutions, namely boson star (BS) type solutions for a complex vector field with mass and nonminimal coupling to the Einstein tensor. Irrespective of the existence of nonminimal coupling, PS solutions are mini-BS type, but the inclusion of it changes properties. For positive nonminimal coupling parameters, PS solutions do not exist for central amplitudes above some certain value due to the singular behavior of the evolution equations. For negative nonminimal coupling parameters, there is no such singular behavior but sufficiently enhanced numerical resolutions are requested for larger amplitudes. Irrespective of the sign of the nonzero nonminimal coupling parameter, PSs with the maximal Arnowitt-Deser-Misner mass and Noether charge are gravitationally bound. Properties of PSs are very similar to those of BSs in scalar-tensor theories including healthy higher-derivative terms. 
\end{abstract}
\pacs{
04.40.-b Self-gravitating systems; continuous media and classical fields in curved spacetime,
04.50.Kd Modified theories of gravity.
}
\date{\today}
\maketitle

%\tableofcontents
%%%%%%%%%%%%%%%%%%%%%%%%%
\section{Introduction}
\label{sec1}

Motivated by the observed acceleration of the Universe,
many modified gravity (MG) models have been proposed.
These models have been tested 
in the various contexts of astrophysics and cosmology
\cite{Clifton:2011jh,Berti:2015itd}.
It is known that 
many MG models
can be expressed 
in terms of the subclasses of the Horndeski theories,
namely the most general scalar-tensor theories with the second-order equations of motion (EOMs)
\cite{Horndeski:1974wa,Deffayet:2009mn,Deffayet:2011gz,Kobayashi:2011nu}.
More recently, 
the extension of the Horndeski theories to the vector-tensor theories
has been explored in Refs. \cite{Tasinato:2014eka,Heisenberg:2014rta,Allys:2015sht,Jimenez:2016isa,DeFelice:2016cri,DeFelice:2016yws,DeFelice:2016uil},
which are called the generalized Proca theories,
as they also correspond to nonlinear extensions of the massive vector field theory
where the $U(1)$ gauge symmetry is explicitly broken.

%%%%%%%%%%%%%%%%%%%%%%%%%%%%%%%% 

Properties of compact objects will be very important to distinguish MG models
in light of future gravitational wave (GW) observations. 
The simplest compact objects are black holes (BHs),
which in the context of generalized Proca theories
have been studied in Refs.
\cite{Chagoya:2016aar,Minamitsuji:2016ydr,Geng:2015kvs,Chagoya:2017fyl,Babichev:2017rti,Heisenberg:2017xda}.
These solutions have exhibited
nontrivial stealth features for the nontrivial Proca and electric charges. 
Neutron star (NS) solutions in a subclass of generalized Proca theories with nonminimal coupling to the Einstein tensor
were studied in Ref. \cite{,Chagoya:2017fyl}.
In this paper, as a candidate of more exotic compact objects,
we investigate boson star (BS) type solutions 
in the generalized complex Proca theory with mass and nonminimal coupling to the Einstein tensor,
namely Proca stars (PSs).
\footnote{In this paper, we will distinguish ``boson stars (BSs)'' and  ``Proca stars (PSs)'' 
for condensates of complex scalar and vector fields, respectively.}

%%%%%%%%%%%%%%%%%%%%%%%%%%%%%%%%

BSs are gravitationally bound nontopological solitons constituted by bosonic particles.
BS solutions have been firstly constructed 
for a complex scalar field with mass $\mu^2 |\phi|^2$ \cite{Kaup:1968zz,Ruffini:1969qy,Friedberg:1986tp,Jetzer:1991jr}.
%%%%
Mass and radius of BSs are typically
$\sim M_p^2/\mu$
and 
$\sim 1/\mu$,
respectively,
where $M_p=\sqrt{\hbar c/G}= 1.221\times 10^{19} {\rm GeV}/c^2$ is the Planck mass.
\footnote{In the rest, we will work in the units of $c=\hbar=1$.}
Thus,
for $\mu\sim 1{\rm GeV}$ close to that of protons and neutrons,
the mass of BSs 
becomes $10^{13}{\rm g}$,
which is much smaller than the Chandrasekhar mass
for their fermionic counterparts
$M_{\rm Ch}\sim M_p^3/\mu^2\gtrsim M_\odot$,
where $M_\odot=1.99\times 10^{33}{\rm g}$ is the Solar mass.
On the other hand, 
for $\mu\sim 10^{-10}{\rm eV}$,
mass and radius of BSs become $\sim M_\odot$ and $\sim 10 {\rm km}$,
respectively,
which may be targets for future GW observations as important as BHs and NSs.
%%%%%%
See Ref. \cite{Schunck:2003kk} and references therein for BS solutions for other scalar potentials.
%%%%%

BSs are characterized by several conserved charges.
The first is the Arnowitt-Deser-Misner (ADM) mass, which corresponds to the gravitational mass. 
The second is the Noether charge $Q$ associated with the global internal symmetry in the scalar field sector,
where $\mu Q$ measures the rest mass energy of bosonic particles.
Thus, the binding energy of a BS is given by $B:=M-\mu Q$.
For $B<0$ a BS is said to be gravitationally bound,
while 
for $B>0$ the excess energy would be translated into kinetic energy of individual bosons.
%%%%%%%%%%
Radial perturbations of BS solutions 
have been studied in Refs. \cite{Gleiser:1988rq,Jetzer:1991jr,Lee:1988av,Gleiser:1988ih}.
%%%%%%%%%%
$Q$ and $M$ are generically the functions of the amplitude of the scalar field at the center $\phi_0$,
i.e., $Q=Q(\phi_0)$ and $M=M(\phi_0)$.
It has been shown that 
the critical solution dividing stable BSs from unstable ones 
satisfies $dQ/d\phi_0 (\phi_{0,c})=dM/d\phi_0 (\phi_{0,c})= 0$ ,
where $\phi_{0,c}$ describes the critical central amplitude
and solutions for $\phi_0>\phi_{0,c}$ are unstable \cite{Gleiser:1988ih,Hawley:2000dt}.
%%% 
Because $B(\phi_{0,c})= M(\phi_{0,c})-\mu Q (\phi_{0,c})<0$,
this critical solution possesses negative binding energy.
In other words,
$B<0$ does not necessarily correspond to perturbative stability of BSs.
%%%%%%%
Nevertheless,
because stable BS solutions satisfy $B<0$,
we may use it as a necessary condition of stability.

In Ref. \cite{Brihaye:2016lin},
BS solutions have been studied
for a complex scalar field
with nonminimal derivative coupling to the Einstein tensor $G^{\mu\nu}\partial_\mu \phi\partial_\nu \bar{\phi}$,
where $\bar \phi$ is the complex conjugate of $\phi$,
which is analogous to a subclass of the Horndeski theories for a real scalar field.
It has been shown that
the inclusion of nonminimal derivative coupling changes properties of BSs.
%%%%%%%%
For a massive complex field,
for positive nonminimal derivative coupling parameters,
the evolution equations to determine the structure of BSs
become singular for central amplitudes above some certain value,
and as a result no BS solutions exist.
On the other hand,
for negative nonminimal derivative coupling parameters,
the evolution equations do not become singular,
but for larger central amplitudes 
enhanced resolutions are requested.
Similar properties have been observed
for BSs in the Einstein-Gauss-Bonnet (EGB) \cite{Hartmann:2013tca,Brihaye:2013zha}
and Einstein dilaton Gauss-Bonnet (EdGB) \cite{Baibhav:2016fot} theories.
%%%%%%%%%%%%%%%%%%

Recently,
BS type solutions for a complex vector field, 
namely PS solutions,
have been studied in Refs. \cite{Brito:2015pxa,Garcia:2016ldc,Brihaye:2017inn}.
For a massive vector field $\mu^2 A^\mu \bar{A}_\mu$,
where $\bar{A}_\mu$ is the complex conjugate of $A_\mu$,
PSs are mini-BS types with masses $\sim M_p^2/\mu$.
The properties of PSs are quite similar to those of BSs,
and 
the critical solution dividing stable and unstable PS solutions
corresponds to that with 
the maximal ADM mass and Noether charge \cite{Brito:2015pxa}.
%%%%%%%
In this paper,
we will investigate PS solutions
in the presence of nonminimal coupling to the Einstein tensor
$G^{\mu\nu}A_\mu \bar{A}_\nu$,
which is analogous to a subclass of the generalized Proca theories
studied in the context of BH physics 
\cite{Chagoya:2016aar,Minamitsuji:2016ydr,Geng:2015kvs,Chagoya:2017fyl,Babichev:2017rti}.
%%%%%%
We will find properties of PSs
which are very similar to those of BSs in the complex scalar-tensor theories
with healthy higher-derivative interactions.
\footnote{By ``healthy,'' 
we mean that there are no ghosty degrees of freedom associated with Ostrogradsky's theorem \cite{Woodard:2015zca}.}

The paper is constructed as follows:
In Sec. \ref{sec2}, 
we will introduce the generalized complex Proca theory with mass and nonminimal coupling to the Einstein tensor
and provide the covariant EOMs.
In Sec. \ref{sec3},
we will arrange EOMs to a set of the evolution equations in the static and spherically symmetric background 
to determine the structure of PSs.
%%%%%
In Sec. \ref{sec4}, we will numerically construct PS solutions and discuss their properties.
Finally, 
Sec. \ref{sec5}
will be devoted to giving a brief summary and conclusion.

%%%%%%%%%%%%%%%%%%%%%%%%%%
\section{The generalized complex Proca theory with mass and nonminimal coupling to the Einstein tensor}
\label{sec2}

We consider the generalized complex Proca theory
with mass and nonminimal coupling to the Einstein tensor
\begin{align}
\label{action}
S&=\int d^4x \sqrt{-g}
\left[
 \frac{1}{2\kappa^2}R
-\frac{1}{4} F^{\mu\nu}{\bar F}_{\mu\nu}
-\frac{1}{2}\left(\mu^2g^{\mu\nu}-\beta G^{\mu\nu}\right)A_\mu {\bar A}_\nu
\right],
\end{align}
where the Greek indices $(\mu,\nu,...)$ run the four-dimensional spacetime,
$g_{\mu\nu}$ is the metric tensor,
$g^{\mu\nu}:=\left(g_{\mu\nu}\right)^{-1}$ is the inverse metric tensor,
$g={\rm det} (g_{\mu\nu})$ is the determinant of $g_{\mu\nu}$,
and 
$R$ and $G_{\mu\nu}$
are the Ricci scalar curvature and the Einstein tensor
associated with $g_{\mu\nu}$.
$A_\mu$ is the complex vector field,
$F_{\mu\nu}:= \partial_\mu A_\nu-\partial_\nu A_\mu$
is the field strength,
quantities with a ``bar''
are the complex conjugates of those without a ``bar'',
$\kappa^2:= 8\pi G$,
$G$ is the gravitational constant, 
$\mu^2$ is mass of the complex vector field,
and $\beta$ measures nonminimal coupling to the Einstein tensor.
In our units, 
the Planck mass $M_p$ is given by $M_p=1/\sqrt{G}$.

We obtain EOMs by varying the action Eq. \eqref{action}
with respect to $g_{\mu\nu}$ and $A_\mu$, respectively.
Varying Eq. \eqref{action} with respect to $g_{\mu\nu}$,
we obtain the gravitational EOM
\begin{align}
\label{grav_eq}
0={\cal E}_{\mu\nu}
&:=T^{(f)}_{\mu\nu}+\beta  T^{(A)}_{\mu\nu}-\frac{1}{\kappa^2}G_{\mu\nu},
\end{align}
where
\begin{align}
\label{tf}
T^{(f)}_{\mu\nu}
&:=
 \frac{1}{2} F_{\mu\rho} {\bar F}_\nu{}^\rho
+\frac{1}{2} {\bar F}_{\mu\rho} F_\nu{}^\rho 
-\frac{1}{4} g_{\mu\nu}F^{\rho\sigma} {\bar F}_{\rho\sigma}
+
\frac{\mu^2}{2}
\left(
   A_\mu {\bar A}_\nu
+ A_\nu {\bar A}_\mu
\right)
-\frac{\mu^2}{2}
g_{\mu\nu} A^\rho {\bar A}_\rho
\end{align}
represents the energy-momentum tensor for the ordinary massive Proca field,
and 
\begin{align}
\label{ta}
T^{(A)}_{\mu\nu} 
&:=\frac{1}{2} 
\left(
   A^\rho {\bar A}_\rho G_{\mu\nu}
+\frac{1}{2} R A_\mu {\bar A}_\nu
+\frac{1}{2} R {\bar A}_\mu A_\nu
\right)
\nonumber\\
&-\frac{1}{2}g_{\mu\nu}
\left(
  \nabla^\rho A_\rho \nabla^\sigma \bA_\sigma
-2\nabla^\rho A^\sigma \nabla_\rho \bA_\sigma
+\nabla^\rho A^\sigma \nabla_\sigma \bA_\rho
-A_\rho \Box \bA^\rho
-\bA_\rho \Box A^\rho
+A^\rho \nabla_\rho \nabla_\sigma \bA^\sigma
+\bA^\rho \nabla_\rho \nabla_\sigma A^\sigma
\right)
\nonumber\\
&-
\left(
  \frac{1}{2}\nabla_\mu A^\rho \nabla_\nu \bA_\rho
 +\frac{1}{2}\nabla_\mu \bA^\rho \nabla_\nu A_\rho
 - \frac{1}{2}\nabla^\rho A_\rho \nabla_{(\mu} \bA_{\nu)}
 - \frac{1}{2}\nabla^\rho \bA_\rho \nabla_{(\mu} A_{\nu)}
\right.
\nonumber\\
&-\frac{1}{2}\nabla_\rho A_{(\mu}\nabla_{\nu)}\bA^\rho
-\frac{1}{2}\nabla_\rho \bA_{(\mu}\nabla_{\nu)} A^\rho
+\frac{1}{2}\nabla_\rho A_{\mu}\nabla^\rho \bA_\nu
+\frac{1}{2}\nabla_\rho A_{\nu}\nabla^\rho \bA_\mu
\nonumber\\
&+\frac{1}{2} A_\rho \nabla_{(\mu} \nabla_{\nu)}\bA^\rho
 +\frac{1}{2} \bA_\rho \nabla_{(\mu} \nabla_{\nu)} A^\rho
 -\frac{1}{2} A^\rho \nabla_{\rho} \nabla_{(\mu}\bA_{\nu)}
 -\frac{1}{2} \bA^\rho \nabla_{\rho} \nabla_{(\mu} A_{\nu)}
\nonumber\\
&\left.
+\frac{1}{2} A_{(\mu} \Box \bA_{\nu)}
 +\frac{1}{2} \bA_{(\mu} \Box A_{\nu)}
-A_{(\mu}\nabla_{\nu)} \nabla_\sigma \bA^\sigma
-\bA_{(\mu}\nabla_{\nu)} \nabla_\sigma A^\sigma
+\frac{1}{2} A_{(\mu|} \nabla_\rho \nabla_{|\nu)} \bA^\rho
+\frac{1}{2} \bA_{(\mu|} \nabla_\rho \nabla_{|\nu)} A^\rho
\right)
\end{align}
represents the effective energy-momentum tensor for nonminimal coupling to the Einstein tensor.
Similarly, 
varying Eq. \eqref{action} with respect to $A_\mu$ and $\bar{A}_\mu$,
we obtain the EOM for the complex vector field  
\begin{align}
\label{vector_eq}
0= {\cal F}_\nu
&:=
 \nabla^\mu F_{\mu\nu}
-\left(
 \mu^2 \delta_{\nu}{}^\mu 
-\beta G_{\nu}{}^\mu 
 \right)
A_\mu,
\end{align}
and its complex conjugate $\overline{\cal F}_\nu=0$,
respectively.
The $U(1)$ gauge symmetry is explicitly broken by mass and nonminimal coupling. 
By taking the derivative of Eq. \eqref{vector_eq},
we obtain
\begin{align}
\label{gauge}
0={\cal G}&:= \mu^2 \nabla^\mu A_\mu-\beta G^{\mu\nu}\nabla_\mu A_\nu,
\end{align}
which generalizes the constraint relation for $\beta=0$,
$\nabla^\mu A_\mu=0$.
In the theory \eqref{action},
there is still global $U(1)$ symmetry 
under $A_\mu\to e^{i \alpha} A_\mu$,
where $\alpha$ is a constant.
The associated Noether current
is given by 
\begin{align}
\label{noether}
j^\mu=\frac{i}{2}
\left(
 \bF^{\mu\nu} A_\nu
-F^{\mu\nu} \bA_\nu
\right),
\end{align}
which satisfies the conservation law $\nabla_\mu  j^\mu=0$.

%%%%%%%%%%%%%%%%%%%%%%%
\section{Proca stars}
\label{sec3}

\subsection{Static and spherically-symmetric spacetime}
\label{sec3a}

In this section,  we discuss PSs in the theory \eqref{action}.
We consider a static and spherically symmetric spacetime
\begin{align}
\label{metric_ansatz}
g_{\mu\nu}dx^\mu dx^\nu
=-\sigma(r)^2 
\left(1-\frac{2m(r)}{r}\right) d{\hat t}^2 
 + 
   \left(1-\frac{2m(r)}{r}\right)^{-1}
dr^2
 + r^2 d\Omega_2^2,
\end{align}
where $\hat t$ and $r$ are the time and radial coordinates, 
$d\Omega_2^2$ is the metric of the unit two sphere,
and 
$m(r)$ and $\sigma(r)$ 
depend only on the radial coordinate $r$.
Correspondingly, 
we consider the ansatz for the vector field \cite{Brito:2015pxa}
\begin{align}
\label{vector_ansatz}
A_\mu dx^\mu
&=e^{-i{\hat\omega} {\hat t}}
 \left(
     a_0(r)d{\hat t}
   +i a_1 (r)dr
 \right),
\end{align}
where $a_0(r)$ and $a_1(r)$ depend only on $r$.
We assume that the frequency ${\hat\omega}$ is real,
ensuring that the vector field neither grows nor decays.
Equation \eqref{vector_ansatz} avoids the restrictions from Derrick's theorem
that forbid stable localized solutions to nonlinear wave equations \cite{Derrick:1964ww}.
The explicit time dependence $e^{-i{\hat\omega} {\hat t}}$
does not induce the time dependence in the metric.
In order to find $m(r)$, $\sigma(r)$, $a_0(r)$, and $a_1(r)$,
we solve the evolution equation derived from EOMs \eqref{grav_eq}, \eqref{vector_eq}, and \eqref{gauge}.

From our ansatz \eqref{metric_ansatz} and \eqref{vector_ansatz},
the $({\hat t}, {\hat t})$, $(r,r)$ and angular components of the gravitational EOM
\eqref{grav_eq}
\begin{align}
\label{grav_eq0}
&
{\cal E}^{\hat t}{}_{\hat t}=0,
\quad 
{\cal E}^r{}_r=0,
\quad
{\cal E}^i{}_i=0,
\end{align}
are given by Eq. \eqref{ecomp} with Eq. \eqref{ecomp2}. 
Similarly,
the $t$ and $r$ components of the vector field EOM
\eqref{vector_eq},
\begin{align}
\label{vector_eq0}
&{\cal F}_{\hat t}=0,
\quad
{\cal F}_r=0,
\end{align}
are given by Eq. \eqref{fcomp} with Eq. \eqref{fcomp2}. 
Finally, 
Eq. \eqref{gauge} is given by Eq. \eqref{gcomp} with Eq. \eqref{gcomp2}. 
These equations are constrained by the identity
\begin{align}
\label{bianchi}
\nabla_\mu 
{\cal E}^\mu{}_r
=
-\frac{1}{2}
  g^{{\hat t}{\hat t}} 
 \left(
{\cal F}_{\hat t} 
\times 
\bar{F}_{{\hat t}r}
 +
\bar{\cal F}_{\hat t}
\times 
F_{{\hat t}r}\right) 
+\frac{1}{2}
\left(
{\cal G} \bA_r
+
 {\bar{\cal  G}} A_r
\right).
\end{align}
%%%%%%%%%%%%%%%%%%%%%%%
Combining them, 
we obtain a set of the evolution equations \eqref{diag1}, \eqref{diag2}, \eqref{diag3}, and \eqref{a1p}
which determine the structure of PSs from the center $r=0$ to spatial infinity $r=\infty$.

%%%%%%%%%%%%%%%%%%%%%%%%%%

%%%%%%%%%%%%%%%%%%%%%%%%%%%%%%
\subsection{Center}
\label{sec3c}

Solving EOMs near the center $r=0$,
we obtain the boundary conditions
\begin{subequations}
\label{bc}
\begin{align}
a_0(r)
&= f_0 
+ \frac{f_0}{6}
\left\{
\left(\mu^2-\frac{{\hat\omega}^2}{\sigma_0^2}\right)
+\frac{f_0^2\kappa^2}{2\sigma_0^2}
\left(\mu^2-\frac{2{\hat\omega}^2}{\sigma_0^2}\right)\beta
\right.
\nonumber
\\
&
\left.
+\frac{f_0^2\kappa^2}
         {36\mu^2\sigma_0^6}
\left(
3f_0^2\mu^2\kappa^2 
\left(3\mu^2\sigma_0^2-14{\hat\omega}^2\right)
+4{\hat\omega}^2
\left(9\mu^2\sigma_0^2-8{\hat\omega}^2\right)
\right)
\beta^2
+O\left(\beta^3\right)
  \right\}
  r^2
+O\left(r^4\right),
\\
a_1(r)
&=-
\frac{f_0 {\hat\omega}}{3\sigma_0^2}
\left\{
1
+\frac{f_0^2\kappa^2}{\sigma_0^2}
 \beta
+\frac{f_0^2\kappa^2}{18\mu^2\sigma_0^4}
\left(
  21 f_0^2\mu^2\kappa^2
-12 \mu^2\sigma_0^2
+16{\hat\omega}^2
\right)
\beta^2
+O\left(\beta^3\right)
\right\}
r
+O\left(r^3\right),
\\
m(r)
&=
 \frac{f_0^2 \kappa^2}{12\sigma_0^2}
\left\{
 \mu^2
+\frac{3f_0^2\mu^2\kappa^2+4{\hat\omega}^2}
        {6\sigma_0^2}
\beta
+
\frac{3f_0^4\mu^2\kappa^4+20f_0^2\kappa^2{\hat\omega}^2}
        {12\sigma_0^4}
\beta^2
+O\left(\beta^3\right)
\right\}
r^3
+O\left(r^5\right),
\\
\sigma(r)
&=\sigma_0
+
\frac{f_0^2 \kappa^2}{\sigma_0}
\left\{
\frac{\mu^2}{4}
+\frac{3f_0^2\mu^2\kappa^2-3\mu^2\sigma_0^2+4{\hat\omega}^2}{18\sigma_0^2}
 \beta
+\frac{f_0^2\kappa^2}{48\sigma_0^4}
\left(
 5f_0^2\mu^2\kappa^2 
-8\mu^2\sigma_0^2
+20{\hat\omega}^2
\right)
\beta^2
+O(\beta^3)
\right\}
r^2
\nonumber\\
&+O\left(r^4\right).
\end{align}
\end{subequations}
With Eq. \eqref{bc},
we numerically integrate
Eqs. \eqref{diag1}, \eqref{diag2}, \eqref{diag3}, and \eqref{a1p}
toward the spatial infinity $r=\infty$.
From Eq. \eqref{bc},  
for $f_0= 0$ we obtain $\sigma(r)=\sigma_0$ and $m(r)=a_0(r)=a_1(r)=0$,
namely the Minkowski solution.

\subsection{Spatial infinity}
\label{sec3d}

For a correct eigenvalue of ${\hat\omega}$,
we find the asymptotically flat solution
where $m(r)$ and $\sigma(r)$ exponentially approach constant values, 
$m_\infty>0$ and $\sigma_\infty>0$,
respectively,
while
$a_0(r)$ and $a_1(r)$ exponentially approach zero as $e^{-\sqrt{\mu^2-{\hat\omega}^2/\sigma_\infty^2}r}$.
Thus, in the large $r$ limit
the metric exponentially approaches
the Schwarzschild form
\begin{align}
\label{metric}
ds^2\to 
-\sigma_\infty^2 
  \left(
   1-\frac{2m_\infty}{r}
  \right)d{\hat t}^2 
+  
\left(1-\frac{2m_\infty}{r}\right)^{-1}
dr^2
+r^2d\Omega_2^2,
\end{align}
where the proper time measured at $r=\infty$ is given by 
$ t=\sigma_\infty {\hat t} $,
and 
correspondingly 
the proper frequency $\omega$ is given by
\begin{align}
\label{hatom}
{\omega}:=\frac{{\hat\omega}}{\sigma_\infty}.
\end{align}
The exponential fall-off condition $e^{-\sqrt{\mu^2-\omega^2}r}$
requires $\omega <\mu$.
We numerically confirmed that 
in the limit $f_0\to 0$, namely in the limit of the Minkowski solution, 
$\omega\to \mu$, leaving no $r$ dependence in the large $r$ limit.

\subsection{ADM mass, Noether charge, effective radius, and compactness}
\label{sec3e}

Having numerical solutions, we then evaluate the conserved quantities which characterize PSs.
The first is associated with the time translational symmetry 
and corresponds to the ADM mass 
\begin{align}
\label{adm}
M:=\frac{m_\infty}{G}=M_p^2 m_\infty.
\end{align}
The second is 
the Noether charge associated with the global $U(1)$ symmetry,
which is given by 
integrating $j^{\hat t}$ in Eq. \eqref{noether}
over a constant time hypersurface 
\begin{align}
\label{charge}
Q
=\int_\Sigma d^3x \sqrt{-g} j^{\hat t}
= 4\pi \int_0^\infty 
dr
\frac{
r^2a_1(r)
\left(
 {\hat\omega} a_1(r)
-a_0'(r)
\right)}
{\sigma(r)}.
\end{align}
As for a BS,
a PS is said to be gravitationally bound if
\begin{align}
\label{sta}
B:=M-\mu Q<0.
\end{align}
As for BSs mentioned in Sec. \ref{sec1},
for PSs
we use Eq. \eqref{sta} as a necessary condition for stability.

We then define the effective radius \cite{Schunck:2003kk} 
\begin{align}
\label{radius}
\cR
&
:=
\frac{1}{Q}
\int_\Sigma d^3x 
\sqrt{-g}
 \left(r j^{\hat t}\right) 
= 
\frac{4\pi}{Q}
 \int_0^\infty 
dr
\frac{
r^3a_1(r)
\left(
 {\hat\omega} a_1(r)
-a_0'(r)
\right)}
{\sigma(r)},
\end{align}
and the (effective) compactness
\begin{align}
\label{compactness}
{\cal C}:=\frac{GM}{\cR}
    =\frac{M}{M_p^2 \cR}
    =\frac{m_\infty}{\cR}.
\end{align}
In contrast to the cases of BHs and NSs,
for PSs $r=\cR$ is not the surface of $A_\mu=0$.
Nevertheless,
due to the exponential falloff  property of $A_\mu$,
$\cR$ represents the characteristic length scale of energy localization.
The parameter ${\cal C}$ is very important to distinguish various compact objects 
in different theories in light of future GW observations.

\subsection{Parameters}
\label{sec3f}

\begin{figure}[h]
\unitlength=1.1mm
\begin{center}
%\begin{picture}(155,40)
  \includegraphics[height=7.5cm,angle=0]{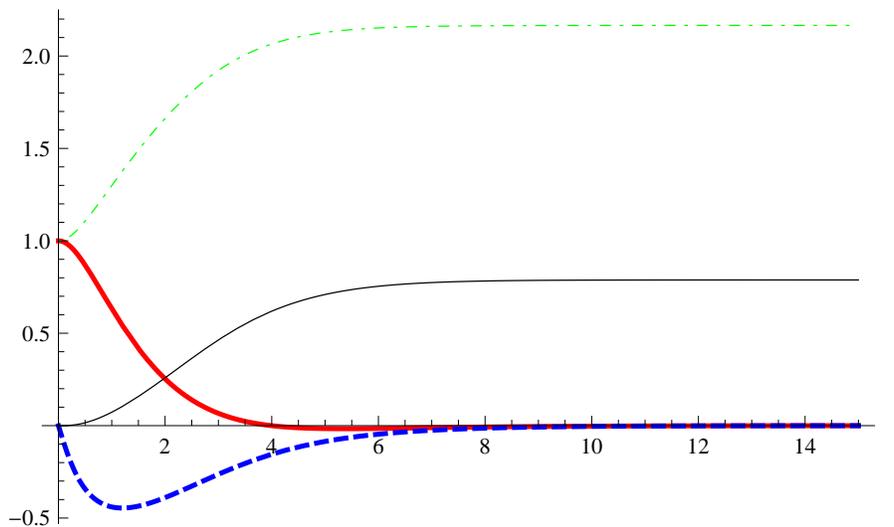}
%\end{picture}
\caption{
$m(r)$, $\sigma(r)$, $a_0(r)$, and $a_1(r)$ are shown as functions of $r$
for $\beta=0.2$ and $f_0=1.0$.
The solid black, green dotted-dashed, thick red, and thick blue dashed curves
correspond to $m(r)$, $\sigma(r)$, $a_0(r)$, and $a_1(r)$, respectively.
Here, we set $\mu=\kappa=\sigma_0=1$.
}
  \label{fig0}
\end{center}
\end{figure}

%%%%%%%
For numerical analyses, we may fix some parameters to unity.
First, by rescaling 
\begin{align}
\label{dimless}
{\hat\omega}\to \frac{{\hat\omega}}{\mu},
\quad
r\to r \mu,
\quad
m(r)\to \mu m(r),
\quad
\sigma(r) \to \sigma(r)
\quad
a_0(r)\to \kappa a_0(r),
\quad
a_1(r)\to \kappa a_1(r),
\end{align} 
we can rewrite Eqs. \eqref{diag1}, \eqref{diag2}, \eqref{diag3}, and \eqref{a1p} 
into equations without $\mu$ and $\kappa$.
Hence, we may work by setting $\mu=\kappa=1$.
%%%%%%
Moreover, 
as $\sigma_0$ corresponds to the time rescaling at the center $r=0$,
we may also set $\sigma_0=1$ for convenience.
In general, then $\sigma_\infty\neq 1$
and quantities measured at the spatial infinity
can be obtained by performing the rescalings discussed in Sec. \ref{sec3d}.
Thus,
the only remaining parameters are $f_0$ and  $\beta$.
For a fixed value of $\beta$,
by varying $f_0$
we numerically integrate them and evaluate the above quantities.
In Fig. \ref{fig0},
by setting $\mu=\kappa=\sigma_0=1$,
$m(r)$, $\sigma(r)$, $a_0(r)$, and $a_1(r)$ are shown as functions of $r$
for $\beta=0.2$ and $f_0=1.0$,
which are very similar to the case of $\beta=0$ \cite{Brito:2015pxa,Garcia:2016ldc}.
%%%%%%%%%%%%%%%% Figure II %%%%%%%%%%%%%%%%
%%%%%%%%%%%%%%%%%%%%%%%%%%%%%%%

Before proceeding, we briefly comment on the massless case $\mu=0$.
In this case, by rescaling 
\begin{align}
r\to {\hat\omega} r ,
\quad 
m\to {\hat\omega} m,
\quad
a_0(r)\to \kappa a_0(r),
\quad 
a_1(r)\to \kappa a_1(r),
\end{align}
the dependence on ${\hat\omega}$ is completely eliminated from EOMs. 
Thus, the problem does not reduce to the eigenvalue one.
As $\beta$ is dimensionless,
for $\mu=0$ there is no physical scale which characterizes PSs.
Thus, there is no PS solution only by nonminimal coupling.

%%%%%%%%%%%%%%%%%%%%%%%%%%%%%%%%
\section{Numerical solutions}
\label{sec4}

For each of $\beta=0.2$, $0.1$, $0$, $-0.1$, and $-0.2$,
for different values of $f_0$
we numerically integrate
the equations \eqref{diag1}, \eqref{diag2}, \eqref{diag3} and \eqref{a1p}
with the boundary conditions \eqref{bc},
and 
find $\omega$ 
that reproduces the asymptotic behaviors discussed in Sec. \ref{sec3d}.
We then evaluate
$M$,  $Q$, $\cR$, and ${\cal C}$
discussed in Sec. \ref{sec3e}.

%%%%%%%%%%%%%%%%%
\subsection{Frequency}
\label{sec4a}

\begin{figure}[h]
\unitlength=1.1mm
\begin{center}
%\begin{picture}(155,40)
  \includegraphics[height=7.5cm,angle=0]{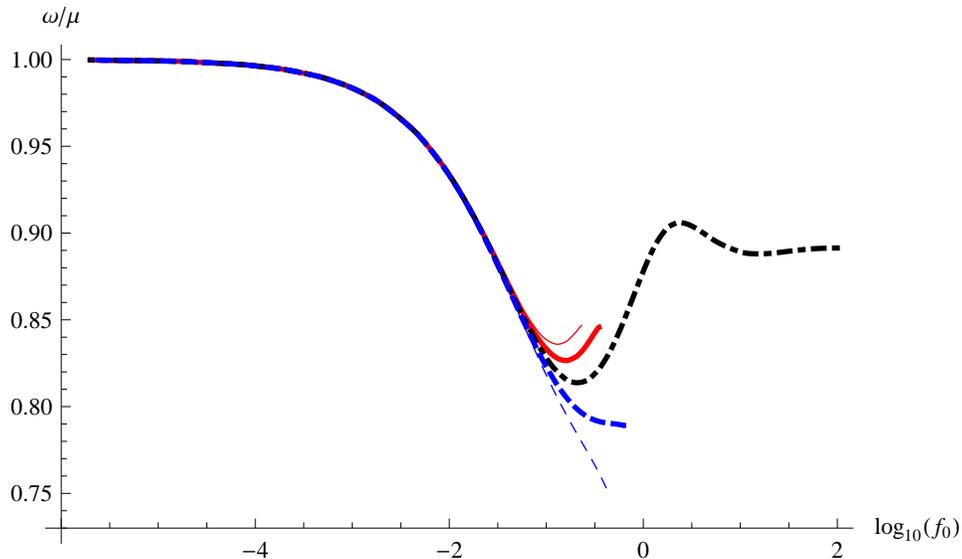}
%\end{picture}
\caption{
$\omega/\mu$
is shown as the function of $f_0$.
The 
solid red,
thick red, 
thick black dotted-dashed,
thick blue dashed,
and 
blue dashed
curves
correspond to
$\beta=0.2$, $0.1$, $0$, $-0.1$, and $-0.2$
respectively.
$f_0$ is shown in $M_p$.
}
  \label{fig1}
\end{center}
\end{figure}

In Fig. \ref{fig1},
for $\beta=0.2$, $0.1$, $0$, $-0.1$, and $-0.2$,
$\omega/\mu$ is shown as a function of $f_0$.
In all cases,
for $f_0\to 0$, $\omega/\mu\to 1$ which reproduces the Minkowski solution.
%%%%%%%%%%%%%%%% Figure II %%%%%%%%%%%%%%%%
%%%%%%%%%%%%%%%%%%%%%%%%%%%%%%%

For $\beta=0$,
there are several branches of PSs for a single value of $\omega$.
As $f_0$ increases from zero,
$\omega/\mu$ decreases from $1$ and reaches the minimal value
$0.814$ for $f_0\approx 0.199 M_p$. 
Then,
$\omega/\mu$ increases
and reaches the local maximal value $0.906$
for $f_0\approx 2.39 M_p$. 
As $f_0$ increases further,
$\omega/\mu$ gradually approaches $0.891$.
%%%%%%%
For nonzero values of $\beta$,
irrespective of its sign,
numerical PS solutions cease to exist 
for $f_0$ above some certain value.
However, the reasons are different for $\beta>0$ and $\beta<0$.

%%%%%%%
For $\beta>0$,
there are several branches of PSs for a single value of $\omega$.
However, for $f_0$ above some certain value depending on $\beta$,
$\tilde{\cal C}_1 $ 
in Eq. \eqref{a1p}
always vanishes at a very small radius, 
and no PS solutions exist.
In our examples,
no PS solutions are found
for $f_0\gtrsim 0.349 M_p$  for $\beta=0.1$
and 
for $f_0\gtrsim  0.231 M_p$ for $\beta=0.2$.
Before reaching these values of $f_0$,
$\omega/\mu$ 
takes the minimal values 
$0.827$ for $f_0\approx 0.160 M_p$ ($\beta=0.1$) 
and 
$0.836$ for $f_0\approx 0.140 M_p$ ($\beta=0.2$).
%%%%%%%%
Thus, for larger nonminimal coupling parameters,
PS solutions cease to exist for smaller amplitudes.
%%%%%%%%

For $\beta<0$, $\omega/\mu$ always decreases.
Although
$\tilde{\cal C}_1$ 
in Eq. \eqref{a1p}
does not vanish at any radius,
it becomes extremely small
for larger values of $f_0$,
making numerical integrations unstable
unless sufficiently high resolutions are taken.
Thus,
although we will show PS solutions only for smaller values of $f_0$
where numerical integration can be performed stably, 
in principle
it does not forbid the existence of PS solutions for larger values of $f_0$.
In our analysis, 
numerical integration could not performed stably
for $f_0\lesssim 0.628 M_p$ for $\beta=-0.1$
and 
for $f_0\lesssim 0.409 M_p$ for $\beta=-0.2$.

\subsection{ADM mass and Noether charge}
\label{sec4b}

\begin{figure}[ht]
\unitlength=1.1mm
\begin{center}
  \includegraphics[height=7.5cm,angle=0]{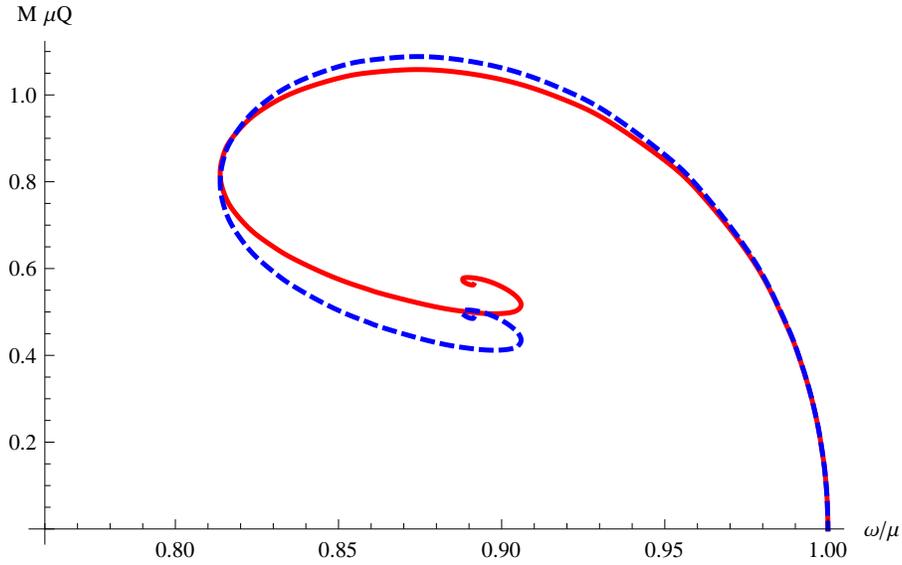}
\caption{
For $\beta=0$,
$M$ and $\mu Q$
are shown
as functions of $\omega/\mu$.
The solid red and blue dashed curves
correspond 
to $M$ and $\mu Q$,
respectively.
Here, 
$M$ and $\mu Q$
are shown in $M_p^2/\mu$.
}
  \label{fig2}
\end{center}
\end{figure}

\begin{figure}[h]
\unitlength=1.1mm
\begin{center}
%\begin{picture}(155,40)
  \includegraphics[height=7.5cm,angle=0]{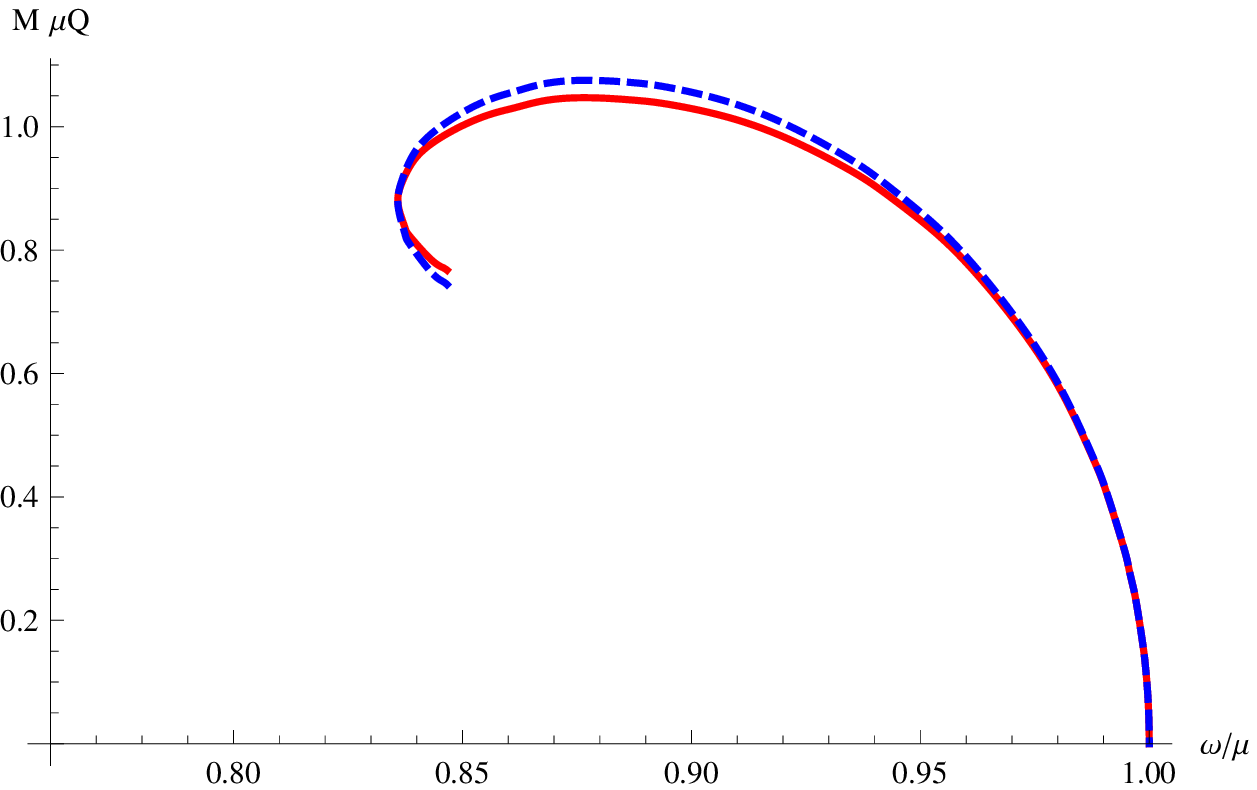}
%\end{picture}
\caption{
For $\beta=0.2$,
$M$ and $\mu Q$
are shown
as functions of $\omega/\mu$.
The solid red and blue dashed curves
correspond 
to  $M$ and $\mu Q$,
respectively.
Here, 
$M$ and $\mu Q$
are shown in $M_p^2/\mu$.
}
  \label{fig3}
\end{center}
\end{figure}

\begin{figure}[h]
\unitlength=1.1mm
\begin{center}
%\begin{picture}(155,40)
  \includegraphics[height=7.5cm,angle=0]{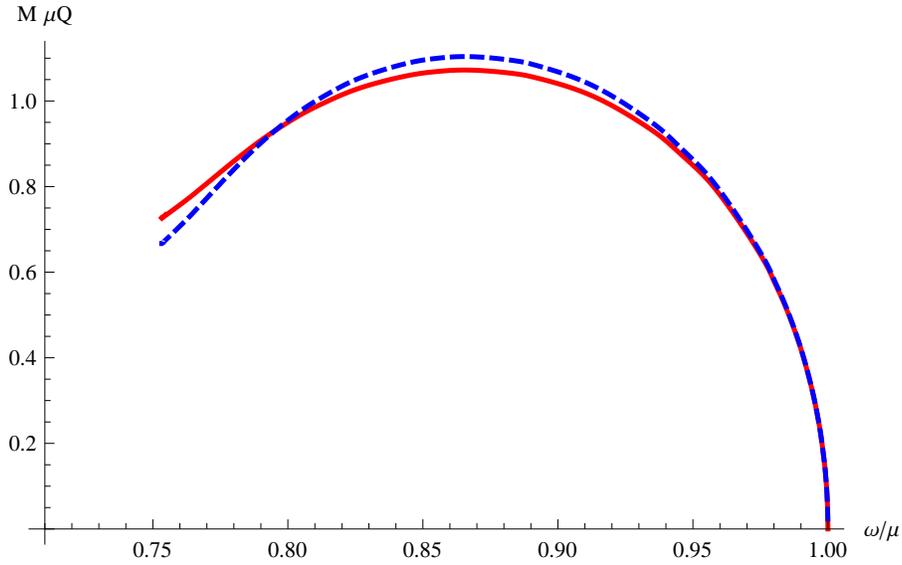}
%\end{picture}
\caption{
For $\beta= -0.2$,
$M$ and $\mu Q$
are shown
as functions of $\omega/\mu$.
The solid red and blue dashed curves
correspond 
to
$M$ and $\mu Q$,
respectively.
Here, 
$M$ and $\mu Q$
are shown in $M_p^2/\mu$.
}
  \label{fig4}
\end{center}
\end{figure}

In Figs. \ref{fig2}-\ref{fig4},
for $\beta=0$, $0.2$, and $-0.2$,
$M$ and $\mu Q$ are shown as functions of $\omega/\mu$.
For all cases,
$f_0=0$ gives the Minkowski solution $M=Q=0$ with $\omega=\mu$.
Moreover, 
even in the presence of the nonminimal coupling to the Einstein tensor 
PS solutions obtained in our analysis are mini-BS type,
where $M$ and $\mu Q$ are typically $\sim M_p^2/\mu$.
%%%%%%%%%%%%%%%% Figure II %%%%%%%%%%%%%%%%
%%%%%%%%%%%%%%%%%%%%%%%%%%%%%%%
\vspace{1.5cm}
%%%%%%%%%%%%%%%% Figure II %%%%%%%%%%%%%%%%
%%%%%%%%%%%%%%%% Figure II %%%%%%%%%%%%%%%%
%%%%%%%%%%%%%%%%%%%%%%%%%%%%%%%

The case of $\beta=0$ is shown in Fig. \ref{fig2}.
As $f_0$ increases from zero,
$M$ and $\mu Q$ increase from zero while $\omega/\mu$ decreases from $1$. 
$M$ and $Q$ then take their maximal values,
$M_{\rm max}\approx 1.06 M_p^2/\mu$   
and 
$\mu Q_{\rm max}\approx 1.09 M_p^2/\mu$  
for $\omega/\mu \approx 0.870$,
which satisfies Eq. \eqref{sta}
and agrees with Ref. \cite{Brito:2015pxa}.
Then 
$\omega/\mu$ still decreases,
while $M$ and $\mu Q$ start to decrease.
After $\omega/\mu$ reaches its minimal value $0.814$,
it starts to increase
while 
$M$ and $\mu Q$ still decrease
until reaching their local minimum values,
$M_{\rm min}\approx  0.496 M_p^2/\mu $  
and 
$\mu Q_{\rm min}\approx  0.412 M_p^2/\mu$,  
respectively.
In this region, Eq. \eqref{sta} is not satisfied.
As $f_0$ increases further,
$M$ and $\mu Q$
eventually converge to
$M\approx 0.563 M_p^2/\mu$ 
and 
$\mu Q\approx 0.486 M_p^2/\mu$,
respectively.

The case of $\beta=0.2$ is shown in Fig. \ref{fig3}.
As $f_0$ increases from zero,
$M$ and $\mu Q$ increase from zero while $\omega/\mu$ decreases from $1$. 
$M$ and $\mu Q$ then take their maximal values,
$M_{\rm max}\approx 1.05M_p^2/\mu $ 
and 
$\mu Q_{\rm max}\approx   1.07 M_p^2/\mu$, 
for $\omega/\mu \approx 0.872$,
which satisfies Eq. \eqref{sta}.
Then as $\omega /\mu$ still decreases,
$M$ and $\mu Q$ start to decrease.
After $\omega/\mu$ reaches its minimal value $0.836$,
$\omega/\mu$ starts to increase
while 
$M$ and $\mu Q$ still decrease.
In this region,
Eq. \eqref{sta} is not satisfied.
As we discussed in Sec. \ref{sec4a},
for $\beta=0.2$,
we can obtain PS solutions up to $f_0=0.231 M_p$,
for which $M\approx 0.768 M_p^2/\mu$ 
and 
$\mu Q\approx 0.744M_p^2/\mu$.

%%%%%%%%%%

Finally, 
the case of $\beta=-0.2$ is shown in Fig. \ref{fig4}.
As $f_0$ increases from zero,
$M$ and $\mu Q$ increase from zero while $\omega/\mu$ decreases from $1$. 
$M$ and $\mu Q$ then take their maximal values,
$M_{\rm max}\simeq  1.07 M_p^2/\mu$  
and 
$\mu Q_{\rm max}\simeq 1.10 M_p^2/\mu$ 
for $\omega/\mu \approx 0.868$,
which satisfies Eq. \eqref{sta}.
As $f_0$ increases further, 
$M$ and $\mu Q$ decrease
while $\omega$ still decreases.
As we discussed in Sec. \ref{sec4a},
for $\beta=-0.2$,
for $f_0$ above $0.409 M_p$,
we could not find numerical PS solutions
due to the problem discussed in Sec. \ref{sec4a}.
For $f_0= 0.409 M_p$,
$M\approx 0.727 M_p^2/\mu$  
and 
$\mu Q\approx 0.667 M_p^2/\mu$.
For $\omega/\mu \lesssim 0.792$,
Eq. \eqref{sta} is not satisfied.

\begin{figure*}[ht]
\unitlength=1.1mm
\begin{center}
\begin{picture}(155,40)
  \includegraphics[height=5.5cm,angle=0]{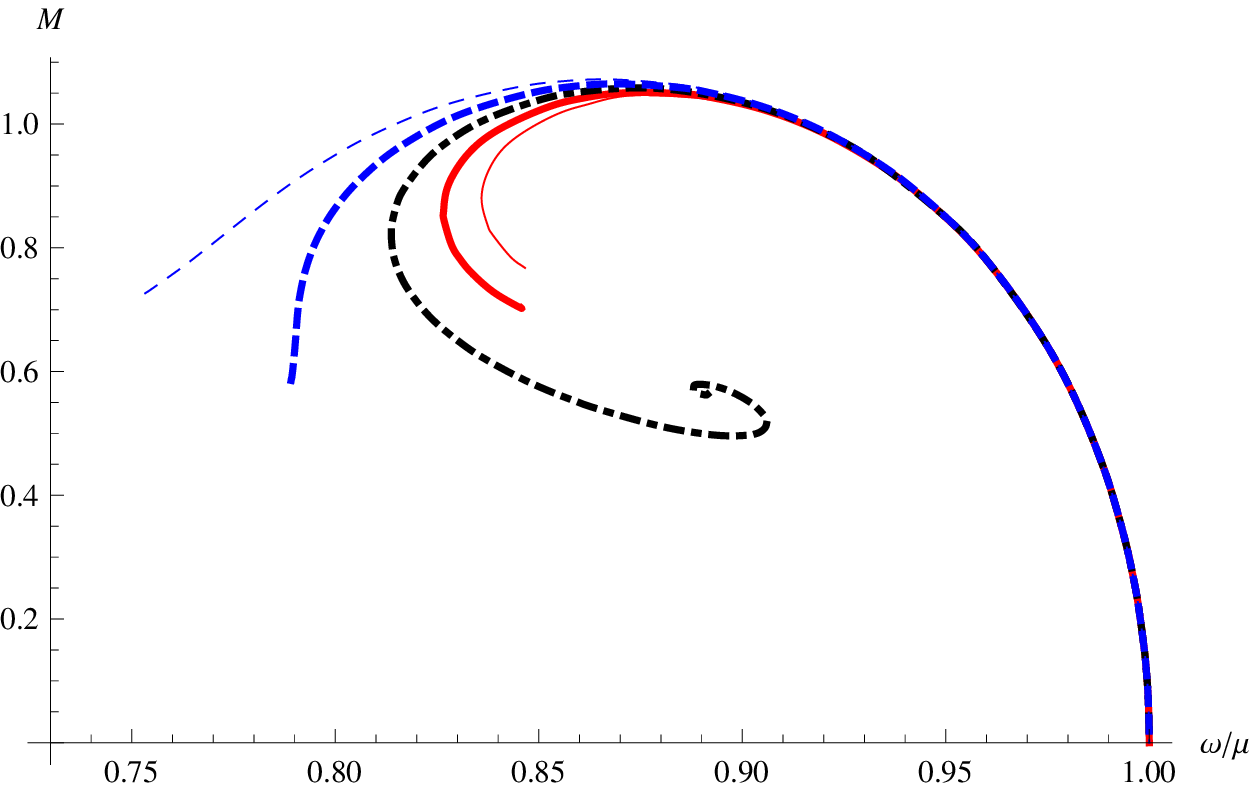}
  \includegraphics[height=5.5cm,angle=0]{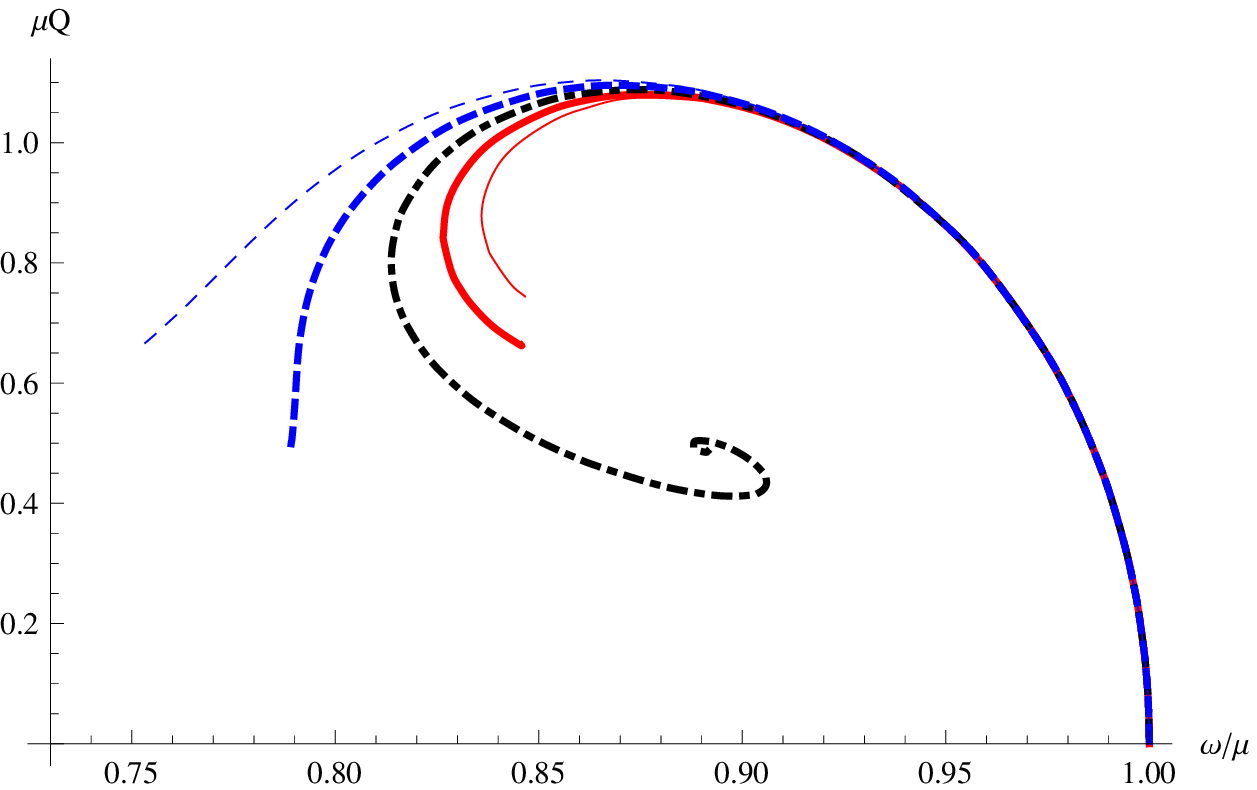}
\end{picture}
\caption{
$M$ (the left panel)
and 
$\mu Q$ (the right panel)
are shown  
as functions of $\omega/\mu$.
In each panel,
the 
solid red, 
thick red, 
black dotted-dashed, 
thick blue dashed,
and blue dashed curves
correspond 
to $\beta=0.2$, $0.1$, $0$, $-0.1$, and $-0.2$,
respectively.
Here, 
$M$ and $\mu Q$
are shown in $M_p^2/\mu$.
}
  \label{fig5}
\end{center}
\end{figure*}

In Fig.~\ref{fig5}, 
$M$ (the left panel) and $\mu Q$ (the right panel)
are shown
for all $\beta=0.2$, $0.1$, $0$, $-0.1$, and $-0.2$.
$M$ and $\mu Q$
almost coincide for larger $\omega/\mu$, i.e., smaller $f_0$.
As $M$ and $\mu Q$ approach their maximal values,
$\beta$ dependence becomes more evident,
and 
for negative (positive) values of $\beta$,
both $M$ and $\mu Q$
take larger (smaller) values than those for $\beta=0$
for the same $\omega$.
%%%%
Moreover, 
for both positive and negative values of $\beta$,
spiraling features observed for $\beta=0$ eventually disappear.
Such behaviors are very similar 
to those observed for BSs 
in the EGB theory \cite{Hartmann:2013tca,Brihaye:2013zha}, the EdGB theory \cite{Baibhav:2016fot} 
and the complex scalar-tensor theory with nonminimal derivative coupling to the Einstein tensor \cite{Brihaye:2016lin}.
Thus, they may be generic for healthy higher-derivative theories. 
\vspace{0.7cm}
%%%%%%%%%%%%%%%% Figure II %%%%%%%%%%%%%%%%
%%%%%%%%%%%%%%%%%%%%%%%%%%%%%%%

%%%%%%%%%%%%%%%%%%%
\subsection{Mass-radius relation}
\label{sec4c}

\begin{figure*}[ht]
\unitlength=1.1mm
\begin{center}
%\begin{picture}(155,40)
  \includegraphics[height=7.5cm,angle=0]{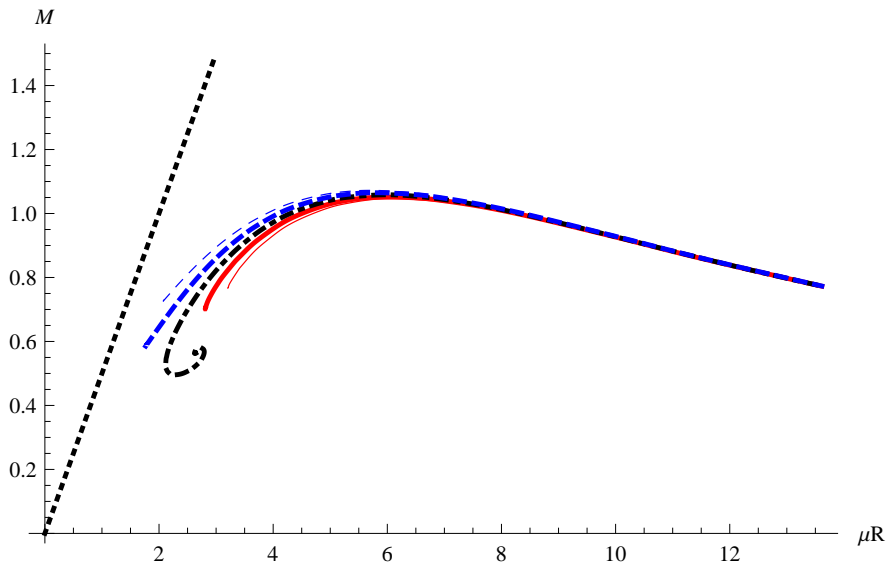}
%\end{picture}
\caption{
$M$ is shown as a function of $\mu \cR$.
The 
solid red,
thick red, 
thick black dotted-dashed,
thick blue dashed,
and 
blue dashed
curves
correspond to
$\beta=0.2$, $0.1$, $0$, $-0.1$, and $-0.2$, respectively,
and the black dotted line represents $\cR=2m_\infty$.
Here, $M$ is shown in $M_p^2/\mu$.
}
  \label{fig6}
\end{center}
\end{figure*}

In Fig. \ref{fig6},
for $\beta=0.2$, $0.1$, $0$, $-0.1$, and $-0.2$,
$M$ is shown
as a function of $\mu \cR$ 
defined in Eq. \eqref{radius}.
In all cases,
for larger $f_0$,
$\cR$ becomes smaller,
which means that energy is localized more efficiently around the center. 
For $\mu \cR\gtrsim 6.0$,
no clear $\beta$ dependence is observed,
while
for $\mu \cR\lesssim 6.0$,
$\beta$ dependence becomes more evident
and 
a larger (smaller) value of $M$
is observed for $\beta<0$ ($\beta>0$). 
For $\beta=0$,
$M$ is a multivalued function of $\cR$,
while
for $\beta\neq 0$, it is a single-valued function of $\cR$.
Very similar behaviors are observed also for $\mu Q$ as a function of $\mu \cR$.
As a reference,
we also show the case of $\cR=2m_\infty$,
although $\cR$ has nothing to do with the Schwarzschild radius of  a BH.  
%%%%%%%%%%%%%%%% Figure II %%%%%%%%%%%%%%%%
%%%%%%%%%%%%%%%%%%%%%%%%%%%%%%%

\begin{figure*}[ht]
\unitlength=1.1mm
\begin{center}
%\begin{picture}(155,40)
  \includegraphics[height=7.5cm,angle=0]{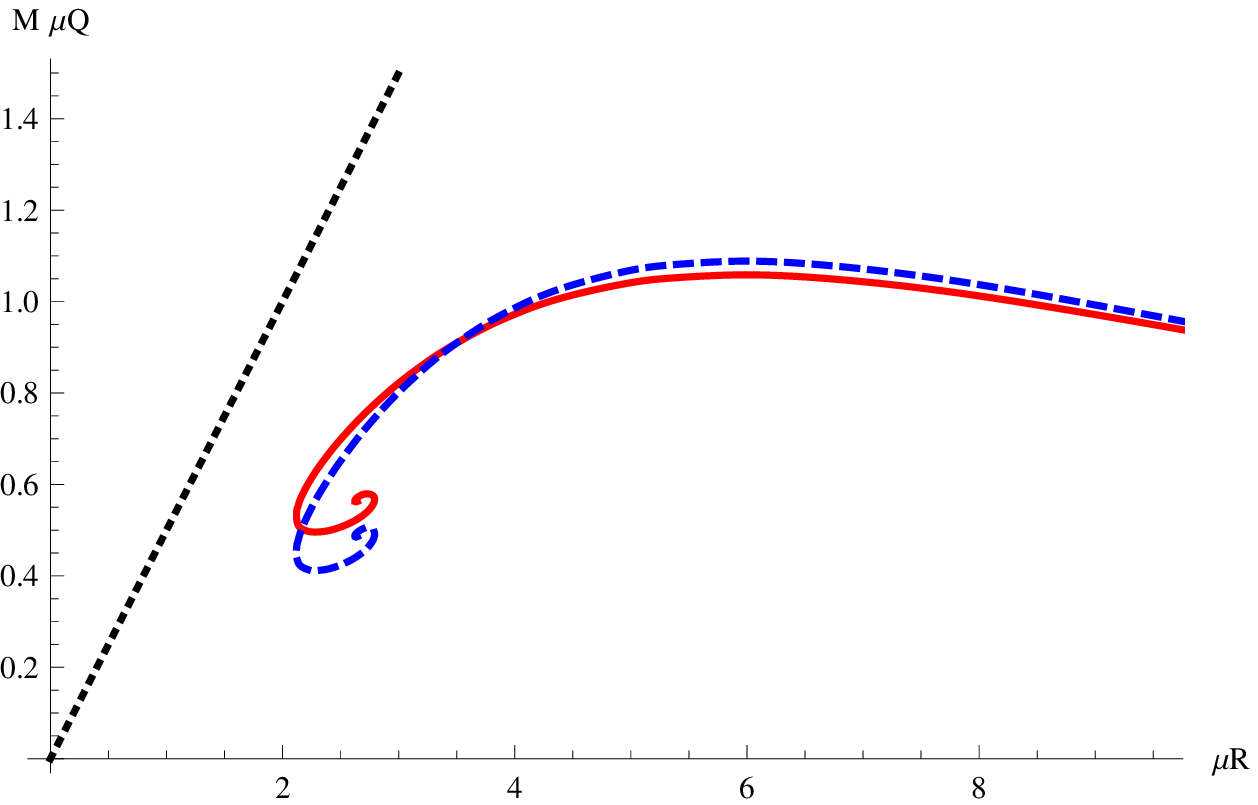}
%\end{picture}
\caption{
$M$ and $\mu Q$
are shown
as functions of $\mu \cR$,
for $\beta=0$.
The solid red and blue dashed curves correspond
to $M$ and $\mu Q$, respectively,
and the black dotted line represents $\cR=2m_\infty$.
Here, $M$ and $\mu Q$ are shown in $M_p^2/\mu$.
}
  \label{fig7}
\end{center}
\end{figure*}

\begin{figure*}[ht]
\unitlength=1.1mm
\begin{center}
%\begin{picture}(155,40)
  \includegraphics[height=7.5cm,angle=0]{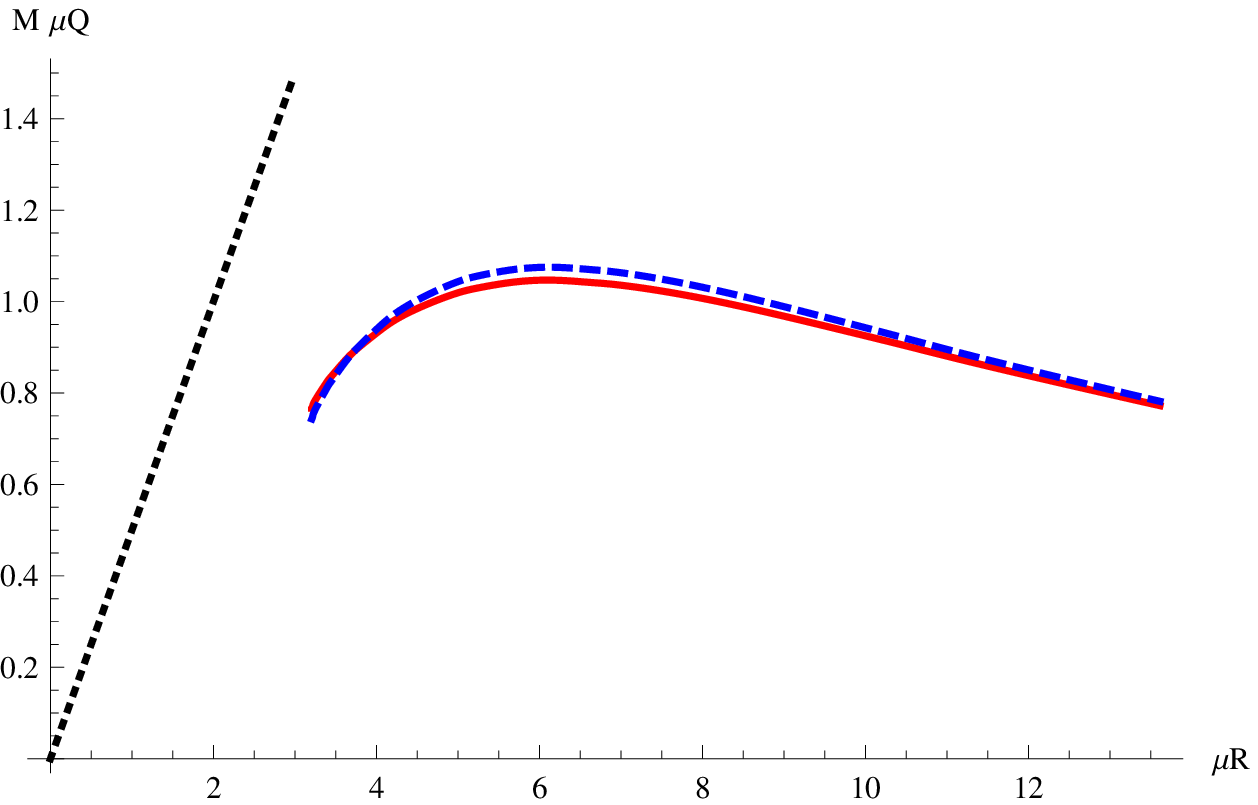}
%\end{picture}
\caption{
$M$ 
and $\mu Q$
are shown
as functions of $\mu \cR$,
for $\beta=0.2$.
The solid red and blue dashed curves correspond
to $M$ and $\mu Q$, respectively,
and the black dotted line represents $\cR=2m_\infty$.
Here, $M$ and $\mu Q$ are shown in $M_p^2/\mu$.
}
  \label{fig8}
\end{center}
\end{figure*}

\begin{figure*}[ht]
\unitlength=1.1mm
\begin{center}
%\begin{picture}(155,40)
  \includegraphics[height=7.5cm,angle=0]{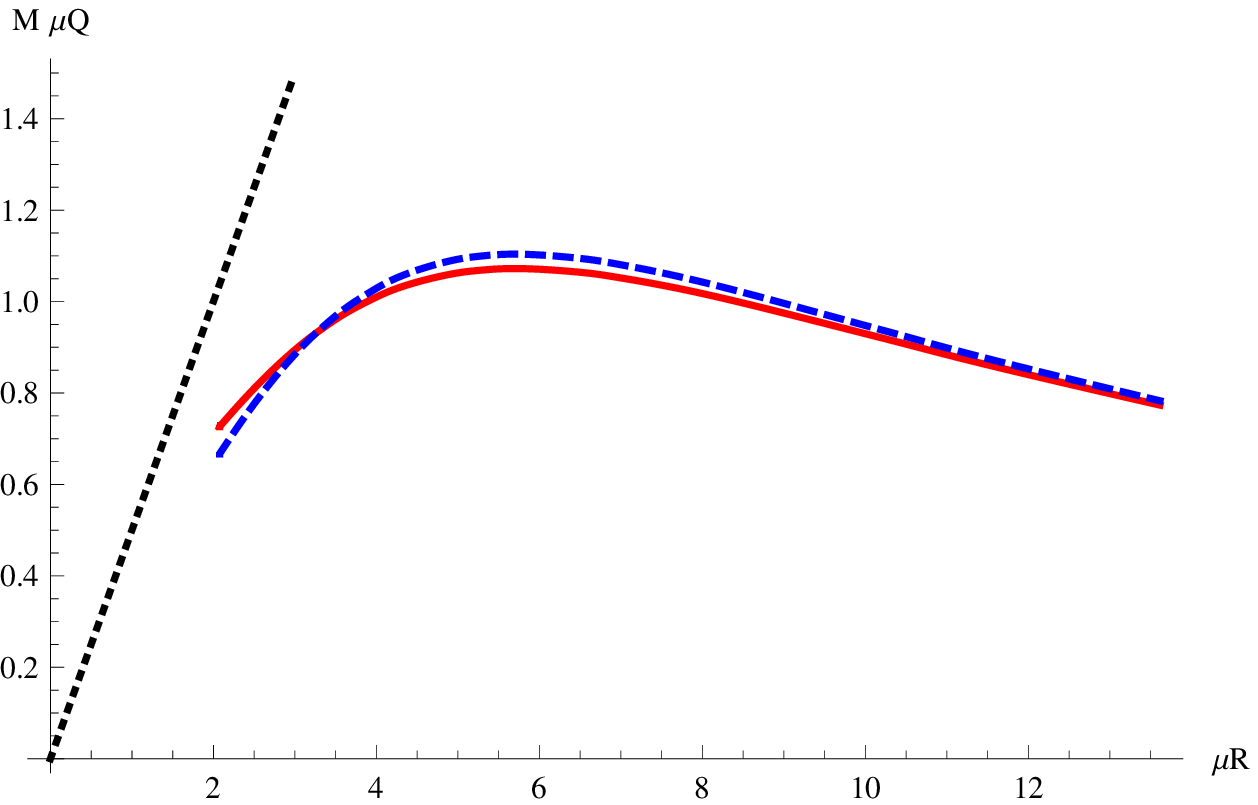}
%\end{picture}
\caption{
$M$ 
and $\mu Q$
are shown
as functions of $\mu \cR$,
for $\beta=-0.2$.
The 
solid red and blue dashed curves correspond
to $M$ and $\mu Q$, respectively,
and the black dotted line represents $\cR=2m_\infty$.
Here, $M$ and $\mu Q$ are shown in $M_p^2/\mu$.
}
  \label{fig9}
\end{center}
\end{figure*}

In Figs. \ref{fig7}-\ref{fig9},
for $\beta=0$, $0.2$, and $-0.2$,
$M$ and $\mu Q$ are shown as functions of $\mu \cR$,
respectively.
For all $\beta=0$, $0.2$, and $-0.2$,
for $\mu \cR\gtrsim 3.5$,
Eq. \eqref{sta}
is satisfied,
while
for $\mu \cR\lesssim 3.5$,
it is not satisfied.
%%%%%%%%%%%%%%%% Figure II %%%%%%%%%%%%%%%%
%%%%%%%%%%%%%%%%%%%%%%%%%%%%%%%
%%%%%%%%%%%%%%%% Figure II %%%%%%%%%%%%%%%%
%%%%%%%%%%%%%%%%%%%%%%%%%%%%%%%
%%%%%%%%%%%%%%%% Figure II %%%%%%%%%%%%%%%%
%%%%%%%%%%%%%%%%%%%%%%%%%%%%%%%

%%%%%%%%%%%%%%%%%%%%%%%%%%%%%%%%%%%%%%
\subsection{Compactness}
\label{sec4d}

\begin{figure*}[ht]
\unitlength=1.1mm
\begin{center}
%\begin{picture}(155,40)
  \includegraphics[height=7.5cm,angle=0]{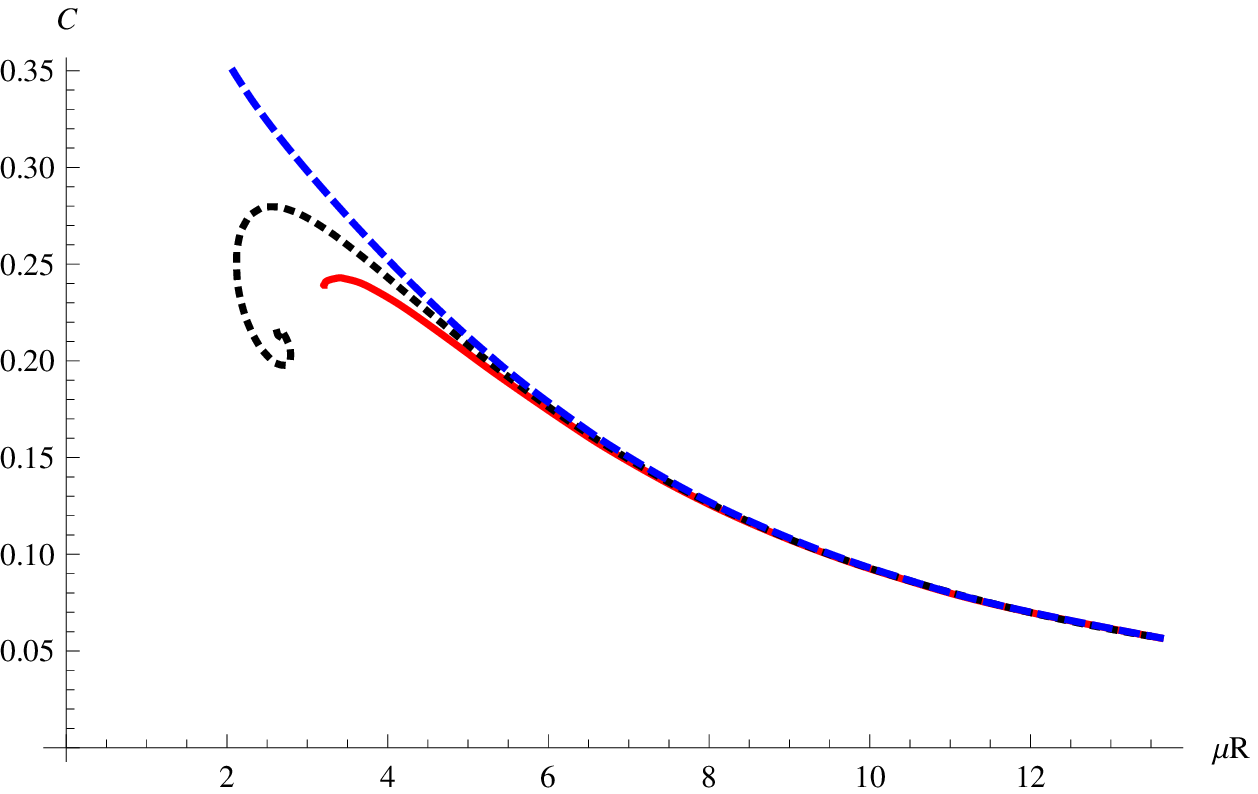}
%\end{picture}
\caption{
${\cal C}$ 
is shown as the function of $\mu \cR$.
The
solid red, 
black dotted,
and blue dashed curves correspond
to
$\beta=0.2$,
$0$,
and 
$-0.2$, respectively.
}
  \label{fig10}
\end{center}
\end{figure*}

In Fig. \ref{fig10},
for $\beta=0.2$, $0$, and $-0.2$,
${\cal C}$ as defined in Eq. \eqref{compactness} 
is shown as the function of $\mu \cR$.
For $\mu \cR\gtrsim 6.0$,
${\cal C}$ becomes smaller
and no clear $\beta$ dependence is observed,
while
for $\mu \cR\lesssim 6.0$,
$\beta$ dependence becomes more evident
and 
a larger (smaller) value of ${\cal C}$ is observed for $\beta<0$ ($\beta>0$).
For $\beta=0$,
$\cal C$ is a multivalued function of $\cR$,
while for $\beta\neq 0$, it is a single-valued function of $\cR$.

For $\beta=0$,
as $f_0$ increases,
$\cR$ decreases,
and 
${\cal C}$ increases and 
takes the maximal value $0.280$ for $\mu \cR\approx 2.50$. 
After reaching the maximal value,
${\cal C}$ starts to decrease,
while $\mu \cR$ still decreases
and 
reaches the minimal value $2.12$.
As $f_0$ further increases,
${\cal C}$ eventually converges to $0.213$.

For $\beta=0.2$,
as $f_0$ increases
while $\mu \cR$ decreases,
${\cal C}$ increases, 
takes the maximal value $0.243$ for $\mu \cR \approx 3.44$,
and then decreases 
until $\mu \cR$ reaches the minimal value $3.21$.

For $\beta=-0.2$,
as $f_0$ increases
while $\mu \cR$ decreases
${\cal C}$ monotonically increases
and takes the maximal value $0.350$
when $\mu \cR$ reaches the minimal value $2.08$
due to the technical problelm discussed in Sec. \ref{sec4a}.
There may be PS solutions for $\mu \cR\lesssim 2.08$,
for which ${\cal C}$ may be larger values than $0.350$.
However, the condition \eqref{sta} is not satisfied for such solutions.

%%%%%%%%%%%%%%%% Figure II %%%%%%%%%%%%%%%%
%%%%%%%%%%%%%%%%%%%%%%%%%%%%%%%

%%%%%%%%%%%%%%%%%%%%%%%%%%%%%%%%
\subsection{Speculation about stability}
\label{sec4e}

Although 
in the given subclass of the generalized complex Proca theory
the explicit analysis of stability will be quite involved,
in this subsection
we will give the speculation about stability
in terms of the insensitivity of the critical central amplitude of the vector field $f_{0,c}$,
for which PSs take the maximal values of the ADM mass $M$ and the Noether charge $Q$,
to the choice of the nonminimal coupling parameter $\beta$.

We have confirmed that
the value of the critical central amplitude,
which in our analysis is found to be $f_{0,c}\approx 0.0399 M_p$,
is insensitive to the choice of $\beta$,
although the absolute maximal values of $M$ and $\mu Q$ both depend on $\beta$.
As we have mentioned in Sec. \ref{sec1},
in the case of $\beta=0$ 
the PS solution with the maximal values of $M$ and $\mu Q$
corresponds to the critical solution 
which divides stable and unstable PS solutions \cite{Brito:2015pxa},
as in the case of BSs in the scalar-tensor theory \cite{Gleiser:1988ih,Hawley:2000dt}.
Thus,
the insensitivity of the value of $f_{0,c}$ to the choice of $\beta$
indicates
that 
even in the presence of nonminimal coupling $\beta \neq 0$
the PS solution with the maximal values of $M$ and $\mu Q$ 
obtained for $f_{0,c}\approx 0.0399 M_p$
would also correspond to the critical PS solution 
which divides stable and unstable PS ones,
and PS solutions for $f_0<f_{0,c}\approx 0.0399 M_p$
would be stable, irrespective of the choice of $\beta$.
The explicit confirmation of this speculation is definitively important,
but will be left for future studies.

%%%%%%%%%%%%%%%%%%%%%%%%%%%%%%%%%%%%%%%%%%
\section{Conclusions}
\label{sec5}

In this paper,
we have investigated boson star type solutions in the generalized complex Proca theory 
with mass and nonminimal coupling to the Einstein tensor,
namely Proca star solutions.
We have numerically constructed PS solutions
for nonzero values of the nonminimal coupling parameter,
and found that the inclusion of it changes properties of PSs.

For positive nonminimal coupling parameters,
PS solutions did not exist for central amplitudes above some certain value,
as the evolution equations to determine the structure of PSs became singular
at very small radii. 
%%%%%
For negative nonminimal coupling parameters,
although there was no such singular behavior,
sufficiently enhanced numerical resolutions were requested for larger amplitudes.
Thus, for larger absolute values of the nonminimal coupling parameter,
spiraling features observed for the vanishing nonminimal coupling 
eventually disappeared.
%%%%%%%%%
Moreover, 
for negative (positive) nonminimal coupling parameters,
for the same frequency 
PSs had larger (smaller) ADM mass and Noether charge
than those for the vanishing nonminimal coupling parameter.
%%%%%%%%%
Irrespective of the sign of the nonzero nonminimal coupling parameter,
PSs with the maximal values of the ADM mass and 
the Noether charge were always gravitationally bound.
%%%%%%%%

Such behaviors of PSs were similar 
to those observed for BSs 
in the EGB theory \cite{Hartmann:2013tca,Brihaye:2013zha}, EdGB theory \cite{Baibhav:2016fot},
and complex scalar-tensor theory with nonminimal derivative coupling to the Einstein tensor \cite{Brihaye:2016lin}.
Thus, they would be generic for gravitational theories including healthy higher-derivative terms. 
Similarities between the scalar-tensor and generalized Proca theories with nonminimal coupling to the Einstein tensor
were pointed out in the context of BH physics
in Refs. \cite{Chagoya:2016aar,Minamitsuji:2016ydr,Chagoya:2017fyl,Babichev:2017rti}.

Finally, 
we have given the speculation on stability of PSs in the given subclass of the generalized complex theory.
We have confirmed that 
the critical central amplitude of the vector field
which provides the PS solutions with the maximal values of the ADM mass and the Noether charge
is insensitive to the choice of the nonminimal coupling parameter.
Since, in the case without nonminimal coupling,
the PS solution with the maximal values of ADM mass and Noether charge
corresponds to the critical PS solution 
dividing stable and unstable PS solutions \cite{Brito:2015pxa},
combined with the confirmed insensitivity to the choice of the nonminimal coupling parameter,
we have speculated that
even in the presence of nonzero nonminimal coupling
PS solutions with the maximal values of ADM mass and Noether charge
correspond to the critical PS solutions dividing stable and unstable PS ones.
The explicit confirmation of this speculation will be left for the future studies.

PS solutions obtained in this paper were mini-BS type.
A definitively interesting question is whether more massive PS solutions can be constructed
in the presence of self-couplings and other nonminiaml couplings to curvatures.
We will leave these issues for our future studies.

%%%%%%%%%%%%%%%%%%%%%%%%
\section*{Acknowledgements}
This work was supported by FCT-Portugal through Grant No. SFRH/BPD/88299/2012. 
We thank Vishal Baibhav for comments.
%%%%%%%%%%%%%%%%%%%%%%%

\appendix

\section{Components of EOMs}
\label{appa}

The $({\hat t},{\hat t})$, $(r,r)$,
and angular components of the gravitational EOM \eqref{grav_eq0}
are given by
\begin{subequations}
\label{ecomp}
\begin{align}
{\cal E}^{\hat t}{}_{\hat t}
&=
-\frac{1}{2r^4\kappa^2 (r-2m(r)) \sigma(r)^2} 
U^{\hat t}{}_{\hat t}(r),
\\
{\cal E}^r{}_r
&=
\frac{1}{2r^3\kappa^2 (r-2m(r)) \sigma(r)^3} U^r{}_r(r),
\\
{\cal E}^i{}_i
&=
\frac{1}{r^4\kappa^2 (r-2m(r))^2 \sigma(r)^4} U^i{}_i (r),
\end{align}
\end{subequations}
where
\begin{subequations}
\label{ecomp2}
\begin{align}
U^{\hat t}{}_{\hat t}(r)
&:=
r^3\kappa^2 a_0(r)^2
\big(\mu^2 r^2+ 2\beta m'(r)\big)
\nonumber \\
&+\big(r-2m (r)\big)
\Big[
 r^4 \kappa^2 a_0'(r)^2
-2r \kappa^2 a_1(r)
 \big(
  r^3{\hat\omega} a_0'(r)
-2\beta \left(r-2m(r)\right)^2\sigma(r)^2 a_1'(r)
 \big)  
\nonumber\\
&
-4r^2\sigma(r)^2 m'(r)
+\kappa^2 a_1(r)^2 
\Big\{
r^4{\hat\omega}^2 
+(r-2m(r))\sigma(r)^2
\big(
\mu^2 r^3+2r\beta +4\beta m(r)-6r\beta m'(r)
\big)
\Big\}
\Big],
\\
U^r{}_r(r)
&=
-4r^2 \beta\kappa^2 a_0(r) 
  \big(r-2m(r)\big)
 \sigma(r)a_0'(r)
+r\kappa^2 a_0(r)^2
\Big(
\sigma(r)
\big(\mu^2 r^3+4\beta m(r)-2r\beta m'(r)\big)
+2r\beta 
 (r-2m(r))\sigma'(r)
\Big)
\nonumber
\\
&-
\big(r-2m(r)\big)
\sigma(r)
\Big[
-2r^3\kappa^2 {\hat\omega} a_1(r)a_0'(r)
+\kappa^2 a_1(r)^2
 \Big\{
 r^3{\hat\omega}^2 
\nonumber\\
&-(r-2m(r))
  \sigma(r)^2
 (\mu^2 r^2 -2\beta+6\beta m'(r)) 
+6\beta (r-2m(r))^2
\sigma(r)\sigma'(r)
 \Big\}
\nonumber\\
&
+r
\Big\{
r^2\kappa^2 a_0'(r)^2
-4\sigma(r)
\big(
  \sigma(r)m'(r)
-(r-2m(r))\sigma'(r)
\big)
\Big\}
\Big],
\\
U^i{}_i(r)
&=
-2r^3\beta \kappa^2
 a_0(r)\big(r-2m(r)\big)\sigma(r)
\nonumber\\
&\times
\Big[
-2r (r-2m(r))a_0'(r)\sigma'(r)
+\sigma(r)
\Big(
a_0'(r) \big(r-3m(r)+rm'(r)\big)
+r (r-2m(r))a_0''(r)
\Big)
\Big]
\nonumber\\
&
+r^2\kappa^2 a_0(r)^2
\Big[
-2r^2\beta (r-2m(r))^2 \sigma'(r)^2
+\sigma(r)^2
\Big\{
  2\beta m(r)^2
-2rm (r) (\mu^2 r^2+\beta -r\beta m''(r))
\nonumber\\
&+r^2(\mu^2 r^2+2\beta m'(r) -2\beta m'(r)^2 -r\beta m''(r))
\Big\}
\nonumber\\
&+r\beta (r-2m(r))\sigma(r)
\Big(
  (r-3m(r)+rm'(r))\sigma'(r)
+ r(r-2m(r))\sigma''(r)
\Big)
\Big]
\nonumber\\
&+ (r-2m(r))^2
  \sigma(r)^2
\Big[
\nonumber\\
&-2r\kappa^2 a_1(r)
\Big(
  r^3{\hat\omega} a_0'(r)
+\beta (r-2m(r)) \sigma(r)a_1'(r)
 \big(-\sigma (r)(m(r)+r (-1+m'(r)))
      +r(r-2m(r))\sigma'(r)
\big)
\Big)
\nonumber\\
&
+\kappa^2 a_1(r)^2
\Big\{
r^4{\hat\omega}^2
+
\sigma(r)^2
\Big(
2\beta m(r)^2
+2rm(r) (\mu^2 r^2 -\beta -r\beta m''(r))
+r^2 (-\mu^2 r^2 +2\beta m'(r)-2\beta m'(r)^2+r\beta m''(r))
\nonumber\\
&
-r\beta (r-2m(r)) \sigma (r)
 ((r+3m(r)-5rm'(r)) \sigma'(r)+r(r-2m(r))\sigma''(r))
\Big)
\Big\}
\nonumber\\
&
+ r^2
\Big\{
r^2(1-2\beta)\kappa^2 a_0'(r)^2
+
2\sigma (r)
\big[
-r(r+m(r)-3rm'(r))\sigma'(r)
+r \big(\sigma(r)m''(r)-(r-2m(r)) \sigma''(r)\big)
\big]
\Big\}
\Big],
\end{align}
\end{subequations}
respectively.

The ${\hat t}$ and $r$ components of the vector field EOM
\eqref{vector_eq0}
are given by
\begin{subequations}
 \label{fcomp}
\begin{align}
{\cal F}_{\hat t}&=\frac{e^{-i{\hat\omega} {\hat t}}}{r^2\sigma(r)} V_{\hat t}(r),
\\
{\cal F}_r&=-\frac{ie^{-i{\hat\omega} {\hat t}} }{r^2(r-2m(r)) \sigma(r)^2} V_r(r),
\end{align}
\end{subequations} 
where
\begin{subequations}
\label{fcomp2}
\begin{align}
V_{\hat t}(r)&
:=-a_0(r) \sigma(r)
\big(\mu^2 r^2+2\beta m'(r)\big)
\nonumber\\
&+(r-2m(r))
\big(
- r a_0'(r)\sigma'(r)
+ a_1(r) \big(-2{\hat\omega}\sigma(r)+r{\hat\omega} \sigma'(r)\big)
+\sigma(r)
(2a_0'(r)-r{\hat\omega} a_1'(r)+ra_0''(r))
\big),
\\
V_r(r)
&:=
r^3{\hat\omega} a_0'(r)
-a_1(r)
\Big(
 r^3{\hat\omega}^2
-(r-2m(r))\sigma(r)^2
(\mu^2 r^2+2\beta m'(r))
+2\beta (r-2m(r))^2\sigma(r)\sigma'(r)
\Big),
\end{align}
\end{subequations}
respectively.

The constraint relation \eqref{gauge} is given by 
\begin{align}
\label{gcomp}
{\cal G}
=\frac{i e^{-i{\hat\omega} {\hat t}}}{r^4(r-2m(r))\sigma(r)^2} W(r),
\end{align}
where
\begin{align}
\label{gcomp2}
W(r)
&:=
 r^3{\hat\omega} a_0(r)(\mu^2r^2+2\beta m'(r))
\nonumber\\
&+ (r-2m(r))\sigma(r)
\Big[
r(r-2m(r))
a_1'(r)
\big(
 \sigma(r) (\mu^2 r^2+2\beta m'(r))
-2\beta (r-2m(r))\sigma'(r)
\big)
\nonumber\\
&+
a_1(r)
\Big\{
-2\sigma(r)
\Big(
 r 
\big(
\mu^2 r^2m'(r)+2\beta m'(r)^2-r(\mu^2 r+\beta m''(r))
\big)
+
m(r)
\big(
-2\beta m'(r)
+r(\mu^2r+2\beta m''(r))
\big)
\Big)
\nonumber\\
&
+(r-2m(r))
\Big(
\big(
\mu^2 r^2-2r\beta-4\beta m(r)+10r\beta m'(r)
\big)
\sigma'(r)
-2r\beta (r-2m(r))\sigma''(r)
\Big)
\Big\}
\Big].
\end{align}
These equations are 
constrained by Eq. \eqref{bianchi}.

\section{Evolution equations}
\label{appb}

Combining ${\cal E}^r{}_r=0$, ${\cal E}^{\hat t}{}_{\hat t}=0$, and ${\cal F}_{\hat t}=0$
in Eqs. \eqref{ecomp} and \eqref{fcomp},
we obtain
the evolution equations
for $m(r)$, $\sigma(r)$, and $a_0'(r)$
as
%%%%%%%%%%%%%%%%%%%%
\begin{subequations}
\label{diag}
\begin{align}
\label{diag1}
m'(r)
&=-\frac{\kappa^2}
      {2r\big[r^2\beta \kappa^2 a_0(r)^2
-\big(2r+3\beta \kappa^2 a_1(r)^2(r-2m(r)) \big)
\big(r-2m(r) \big)
\sigma(r)^2
\big]}
\nonumber\\
&
\times 
\Big\{
\mu^2 r^5 a_0(r)^2
+\big(r-2m(r)\big)
\Big[
 a_1(r)^2 
\big(
 r^4{\hat\omega}^2
+ \big(r-2m(r)\big)
  \big(\mu^2r^3+2r\beta+4\beta m(r)\big)
\sigma(r)^2
\big)
\nonumber\\
&
+r^4 a_0'(r)^2
-2r a_1(r)
\big(
 r^3{\hat\omega} a_0'(r)
-2\beta (r-2m(r))^2\sigma(r)^2a_1'(r)
\big)
\Big]
\Big\},
\\
\label{diag2}
\sigma'(r)
&=-\frac{\kappa^2\sigma(r)}
         {r\big(r-2m(r)\big)\big[r^2\beta \kappa^2 a_0(r)^2
-\big(2r+3\beta \kappa^2 a_1(r)^2(r-2m(r)) \big)
\big(r-2m(r)\big)\sigma(r)^2
\big]}
\nonumber\\
&\times
\Big\{
a_0(r)^2
\big(
\mu^2 r^5
+2r^2\beta m(r)
\big)
-2r^3\beta a_0(r) 
 \big(r-2m(r)\big)a_0'(r)
\nonumber\\
&+a_1(r) \big(r-2m(r)\big)^2
 \sigma(r)^2
\big[
a_1(r) \big(\mu^2 r^3+2\beta m(r)\big)
+2r\beta
 \big(r-2m(r)\big)
 a_1'(r)
\big]
\Big\},
\\
\label{diag3}
a_0''(r)
&= \frac{1}{r(r-2m(r))}
\times
\Big\{
 \mu^2 r^2 a_0(r)
+2r{\hat\omega} a_1(r)
-4{\hat\omega} a_1(r)m(r)
-2r a_0'(r)
+4m(r) a_0'(r)
\nonumber\\
&
+
\frac{r\kappa^2 \big({\hat\omega} a_1(r)-a_0'(r)\big)}
      {2r^2\beta \kappa^2 a_0(r)^2
     -2\big(2r +3\beta\kappa^2 a_1(r)^2 (r-2m(r))
        \big)
      (r-2m(r))
      \sigma(r)^2}
\nonumber\\
&\times
\Big[
  a_0(r)^2 \big(\mu^2 r^4+4r\beta m(r)\big)
-4r^2\beta a_0(r) \big(r-2m(r)\big)a_0'(r)
\nonumber\\
&-\big(r-2m(r)\big)
\Big(
a_1(r)^2
\big(
 r^3{\hat\omega}^2
-(\mu^2r^2-2\beta) (r-2m(r))\sigma(r)^2
\big)
-2r^3{\hat\omega}
  a_1(r) a_0'(r)
+r^3 a_0'(r)^2
\Big)
\Big]
\nonumber\\
&+ r^2{\hat\omega} a_1'(r)
-2r{\hat\omega} m(r)a_1'(r)
\nonumber\\
&+
\frac{\kappa^2\big(-2\beta a_0(r)+r{\hat\omega} a_1(r)-ra_0'(r)\big)}
 {2r
\Big(
  r^2\beta\kappa^2a_0(r)^2
-\big(
  2r
+3\beta \kappa^2 a_1(r)^2 (r-2m(r))
  \big)
  \big(r-2m(r)\big) 
\sigma(r)^2
  \Big)}
\nonumber\\
&\times
\Big[
  \mu^2 r^5 a_0(r)^2
+ \big(r-2m(r)\big)
\Big(
a_1(r)^2
\big(
r^4{\hat\omega}^2
+(r-2m(r))
 \big(
  \mu^2 r^3+2r\beta+4\beta m(r)
 \big)
\sigma(r)^2
\big)
\nonumber\\
&
+r^4a_0'(r)^2
-2r a_1(r)
\big(
 r^3{\hat\omega} a_0'(r)
-2\beta (r-2m(r))^2\sigma(r)^2a_1'(r)
\big)
\Big)
\Big]
\Big\}.
\end{align}
\end{subequations}
%%%%%%%%%%%%%%%%%%%%%%%%%%

Then, 
${\cal E}^i{}_i=0$ in Eq. \eqref{ecomp}
and
${\cal G}=0$ \eqref{gcomp}
contain the combination
\begin{align}
m''(r)
-\left(r-2m(r)\right)
 \frac{\sigma''(r)}{\sigma(r)}.
\end{align}
By combining them,
we obtain 
\begin{align}
\label{cons}
a_0''(r)
={\cal C}
\left[
a_0(r),
a_1(r), 
a_0'(r),
a'_1(r), 
m(r), 
\sigma(r),
m'(r),
\sigma'(r)
%,a_0''(r)
\right],
\end{align}
where ${\cal C}$ is the nonlinear combination of the given variables,
which is too involved to be shown explicitly.
Then substituting Eqs. \eqref{diag} into Eq. \eqref{cons},
we can replace $m'(r)$, $\sigma'(r)$, and $a_0''(r)$ in Eq. \eqref{cons}
with the lower derivative terms,
and obtain the equation
\begin{align}
\label{c2}
\tilde{\cal C}_0 
\left[
a_0(r),a_1(r), a_0'(r), m(r), \sigma(r)
\right]
+
\tilde{\cal C}_1 
\left[
a_0(r),a_1(r), a_0'(r), m(r), \sigma(r)
\right]
a_1'(r)
=0,
\end{align}
where 
${\tilde C}_i$ ($i=0,1$)
are the nonlinear combinations
of the given variables,
which are too involved to be shown explicitly.
Hence, we can solve Eq. \eqref{c2} 
in terms of $a_1'(r)$ as
\begin{align}
\label{a1p}
a_1'(r)
&=
-\frac{\tilde{\cal C}_0 
\left[
a_0(r),a_1(r), a_0'(r), m(r), \sigma(r)
\right]}
         {\tilde{\cal C}_1 
\left[
a_0(r),a_1(r), a_0'(r), m(r), \sigma(r)
\right]}.
\end{align}
BS solutions do not exist
if $\tilde{\cal C}_1 
\left[
a_0(r),a_1(r), a_0'(r), m(r), \sigma(r)
\right]$ vanishes at some radius.

%%%%%%%%%%%%%%%%%%%%%%%%%%%%%%%%%%%%%%%%%%%
\bibliography{bibmonster} 

\begin{thebibliography}{37}%
\makeatletter
\providecommand \@ifxundefined [1]{%
 \@ifx{#1\undefined}
}%
\providecommand \@ifnum [1]{%
 \ifnum #1\expandafter \@firstoftwo
 \else \expandafter \@secondoftwo
 \fi
}%
\providecommand \@ifx [1]{%
 \ifx #1\expandafter \@firstoftwo
 \else \expandafter \@secondoftwo
 \fi
}%
\providecommand \natexlab [1]{#1}%
\providecommand \enquote  [1]{``#1''}%
\providecommand \bibnamefont  [1]{#1}%
\providecommand \bibfnamefont [1]{#1}%
\providecommand \citenamefont [1]{#1}%
\providecommand \href@noop [0]{\@secondoftwo}%
\providecommand \href [0]{\begingroup \@sanitize@url \@href}%
\providecommand \@href[1]{\@@startlink{#1}\@@href}%
\providecommand \@@href[1]{\endgroup#1\@@endlink}%
\providecommand \@sanitize@url [0]{\catcode `\\12\catcode `\$12\catcode
  `\&12\catcode `\#12\catcode `\^12\catcode `\_12\catcode `\%12\relax}%
\providecommand \@@startlink[1]{}%
\providecommand \@@endlink[0]{}%
\providecommand \url  [0]{\begingroup\@sanitize@url \@url }%
\providecommand \@url [1]{\endgroup\@href {#1}{\urlprefix }}%
\providecommand \urlprefix  [0]{URL }%
\providecommand \Eprint [0]{\href }%
\providecommand \doibase [0]{http://dx.doi.org/}%
\providecommand \selectlanguage [0]{\@gobble}%
\providecommand \bibinfo  [0]{\@secondoftwo}%
\providecommand \bibfield  [0]{\@secondoftwo}%
\providecommand \translation [1]{[#1]}%
\providecommand \BibitemOpen [0]{}%
\providecommand \bibitemStop [0]{}%
\providecommand \bibitemNoStop [0]{.\EOS\space}%
\providecommand \EOS [0]{\spacefactor3000\relax}%
\providecommand \BibitemShut  [1]{\csname bibitem#1\endcsname}%
\let\auto@bib@innerbib\@empty
%</preamble>
\bibitem [{\citenamefont {Clifton}\ \emph {et~al.}(2012)\citenamefont
  {Clifton}, \citenamefont {Ferreira}, \citenamefont {Padilla},\ and\
  \citenamefont {Skordis}}]{Clifton:2011jh}%
  \BibitemOpen
  \bibfield  {author} {\bibinfo {author} {\bibfnamefont {Timothy}\ \bibnamefont
  {Clifton}}, \bibinfo {author} {\bibfnamefont {Pedro~G.}\ \bibnamefont
  {Ferreira}}, \bibinfo {author} {\bibfnamefont {Antonio}\ \bibnamefont
  {Padilla}}, \ and\ \bibinfo {author} {\bibfnamefont {Constantinos}\
  \bibnamefont {Skordis}},\ }\bibfield  {title} {\enquote {\bibinfo {title}
  {{Modified Gravity and Cosmology}},}\ }\href {\doibase
  10.1016/j.physrep.2012.01.001} {\bibfield  {journal} {\bibinfo  {journal}
  {Phys.Rept.}\ }\textbf {\bibinfo {volume} {513}},\ \bibinfo {pages} {1--189}
  (\bibinfo {year} {2012})},\ \Eprint {http://arxiv.org/abs/1106.2476}
  {arXiv:1106.2476 [astro-ph.CO]} \BibitemShut {NoStop}%
%%CITATION = ARXIV:1106.2476;%%
\bibitem [{\citenamefont {Berti}\ \emph {et~al.}(2015)\citenamefont {Berti}
  \emph {et~al.}}]{Berti:2015itd}%
  \BibitemOpen
  \bibfield  {author} {\bibinfo {author} {\bibfnamefont {Emanuele}\
  \bibnamefont {Berti}} \emph {et~al.},\ }\bibfield  {title} {\enquote
  {\bibinfo {title} {{Testing General Relativity with Present and Future
  Astrophysical Observations}},}\ }\href {\doibase
  10.1088/0264-9381/32/24/243001} {\bibfield  {journal} {\bibinfo  {journal}
  {Class. Quant. Grav.}\ }\textbf {\bibinfo {volume} {32}},\ \bibinfo {pages}
  {243001} (\bibinfo {year} {2015})},\ \Eprint
  {http://arxiv.org/abs/1501.07274} {arXiv:1501.07274 [gr-qc]} \BibitemShut
  {NoStop}%
%%CITATION = ARXIV:1501.07274;%%
\bibitem [{\citenamefont {Horndeski}(1974)}]{Horndeski:1974wa}%
  \BibitemOpen
  \bibfield  {author} {\bibinfo {author} {\bibfnamefont {Gregory~Walter}\
  \bibnamefont {Horndeski}},\ }\bibfield  {title} {\enquote {\bibinfo {title}
  {{Second-order scalar-tensor field equations in a four-dimensional space}},}\
  }\href {\doibase 10.1007/BF01807638} {\bibfield  {journal} {\bibinfo
  {journal} {Int.J.Theor.Phys.}\ }\textbf {\bibinfo {volume} {10}},\ \bibinfo
  {pages} {363--384} (\bibinfo {year} {1974})}\BibitemShut {NoStop}%
%%CITATION = IJTPB,10,363;%%
\bibitem [{\citenamefont {Deffayet}\ \emph {et~al.}(2009)\citenamefont
  {Deffayet}, \citenamefont {Deser},\ and\ \citenamefont
  {Esposito-Far\`ese}}]{Deffayet:2009mn}%
  \BibitemOpen
  \bibfield  {author} {\bibinfo {author} {\bibfnamefont {C.}~\bibnamefont
  {Deffayet}}, \bibinfo {author} {\bibfnamefont {S.}~\bibnamefont {Deser}}, \
  and\ \bibinfo {author} {\bibfnamefont {G.}~\bibnamefont
  {Esposito-Far\`ese}},\ }\bibfield  {title} {\enquote {\bibinfo {title}
  {{Generalized Galileons: All scalar models whose curved background extensions
  maintain second-order field equations and stress-tensors}},}\ }\href
  {\doibase 10.1103/PhysRevD.80.064015} {\bibfield  {journal} {\bibinfo
  {journal} {Phys. Rev.}\ }\textbf {\bibinfo {volume} {D80}},\ \bibinfo {pages}
  {064015} (\bibinfo {year} {2009})},\ \Eprint {http://arxiv.org/abs/0906.1967}
  {arXiv:0906.1967 [gr-qc]} \BibitemShut {NoStop}%
%%CITATION = ARXIV:0906.1967;%%
\bibitem [{\citenamefont {Deffayet}\ \emph {et~al.}(2011)\citenamefont
  {Deffayet}, \citenamefont {Gao}, \citenamefont {Steer},\ and\ \citenamefont
  {Zahariade}}]{Deffayet:2011gz}%
  \BibitemOpen
  \bibfield  {author} {\bibinfo {author} {\bibfnamefont {C.}~\bibnamefont
  {Deffayet}}, \bibinfo {author} {\bibfnamefont {Xian}\ \bibnamefont {Gao}},
  \bibinfo {author} {\bibfnamefont {D.A.}\ \bibnamefont {Steer}}, \ and\
  \bibinfo {author} {\bibfnamefont {G.}~\bibnamefont {Zahariade}},\ }\bibfield
  {title} {\enquote {\bibinfo {title} {{From k-essence to generalised
  Galileons}},}\ }\href {\doibase 10.1103/PhysRevD.84.064039} {\bibfield
  {journal} {\bibinfo  {journal} {Phys.Rev.}\ }\textbf {\bibinfo {volume}
  {D84}},\ \bibinfo {pages} {064039} (\bibinfo {year} {2011})},\ \Eprint
  {http://arxiv.org/abs/1103.3260} {arXiv:1103.3260 [hep-th]} \BibitemShut
  {NoStop}%
%%CITATION = ARXIV:1103.3260;%%
\bibitem [{\citenamefont {Kobayashi}\ \emph {et~al.}(2011)\citenamefont
  {Kobayashi}, \citenamefont {Yamaguchi},\ and\ \citenamefont
  {Yokoyama}}]{Kobayashi:2011nu}%
  \BibitemOpen
  \bibfield  {author} {\bibinfo {author} {\bibfnamefont {Tsutomu}\ \bibnamefont
  {Kobayashi}}, \bibinfo {author} {\bibfnamefont {Masahide}\ \bibnamefont
  {Yamaguchi}}, \ and\ \bibinfo {author} {\bibfnamefont {Jun'ichi}\
  \bibnamefont {Yokoyama}},\ }\bibfield  {title} {\enquote {\bibinfo {title}
  {{Generalized G-inflation: Inflation with the most general second-order field
  equations}},}\ }\href {\doibase 10.1143/PTP.126.511} {\bibfield  {journal}
  {\bibinfo  {journal} {Prog. Theor. Phys.}\ }\textbf {\bibinfo {volume}
  {126}},\ \bibinfo {pages} {511--529} (\bibinfo {year} {2011})},\ \Eprint
  {http://arxiv.org/abs/1105.5723} {arXiv:1105.5723 [hep-th]} \BibitemShut
  {NoStop}%
%%CITATION = ARXIV:1105.5723;%%
\bibitem [{\citenamefont {Tasinato}(2014)}]{Tasinato:2014eka}%
  \BibitemOpen
  \bibfield  {author} {\bibinfo {author} {\bibfnamefont {Gianmassimo}\
  \bibnamefont {Tasinato}},\ }\bibfield  {title} {\enquote {\bibinfo {title}
  {{Cosmic Acceleration from Abelian Symmetry Breaking}},}\ }\href {\doibase
  10.1007/JHEP04(2014)067} {\bibfield  {journal} {\bibinfo  {journal} {JHEP}\
  }\textbf {\bibinfo {volume} {04}},\ \bibinfo {pages} {067} (\bibinfo {year}
  {2014})},\ \Eprint {http://arxiv.org/abs/1402.6450} {arXiv:1402.6450
  [hep-th]} \BibitemShut {NoStop}%
%%CITATION = ARXIV:1402.6450;%%
\bibitem [{\citenamefont {Heisenberg}(2014)}]{Heisenberg:2014rta}%
  \BibitemOpen
  \bibfield  {author} {\bibinfo {author} {\bibfnamefont {Lavinia}\ \bibnamefont
  {Heisenberg}},\ }\bibfield  {title} {\enquote {\bibinfo {title}
  {{Generalization of the Proca Action}},}\ }\href {\doibase
  10.1088/1475-7516/2014/05/015} {\bibfield  {journal} {\bibinfo  {journal}
  {JCAP}\ }\textbf {\bibinfo {volume} {1405}},\ \bibinfo {pages} {015}
  (\bibinfo {year} {2014})},\ \Eprint {http://arxiv.org/abs/1402.7026}
  {arXiv:1402.7026 [hep-th]} \BibitemShut {NoStop}%
%%CITATION = ARXIV:1402.7026;%%
\bibitem [{\citenamefont {Allys}\ \emph {et~al.}(2016)\citenamefont {Allys},
  \citenamefont {Peter},\ and\ \citenamefont {Rodriguez}}]{Allys:2015sht}%
  \BibitemOpen
  \bibfield  {author} {\bibinfo {author} {\bibfnamefont {Erwan}\ \bibnamefont
  {Allys}}, \bibinfo {author} {\bibfnamefont {Patrick}\ \bibnamefont {Peter}},
  \ and\ \bibinfo {author} {\bibfnamefont {Yeinzon}\ \bibnamefont
  {Rodriguez}},\ }\bibfield  {title} {\enquote {\bibinfo {title} {{Generalized
  Proca action for an Abelian vector field}},}\ }\href {\doibase
  10.1088/1475-7516/2016/02/004} {\bibfield  {journal} {\bibinfo  {journal}
  {JCAP}\ }\textbf {\bibinfo {volume} {1602}},\ \bibinfo {pages} {004}
  (\bibinfo {year} {2016})},\ \Eprint {http://arxiv.org/abs/1511.03101}
  {arXiv:1511.03101 [hep-th]} \BibitemShut {NoStop}%
%%CITATION = ARXIV:1511.03101;%%
\bibitem [{\citenamefont {Beltran~Jimenez}\ and\ \citenamefont
  {Heisenberg}(2016)}]{Jimenez:2016isa}%
  \BibitemOpen
  \bibfield  {author} {\bibinfo {author} {\bibfnamefont {Jose}\ \bibnamefont
  {Beltran~Jimenez}}\ and\ \bibinfo {author} {\bibfnamefont {Lavinia}\
  \bibnamefont {Heisenberg}},\ }\bibfield  {title} {\enquote {\bibinfo {title}
  {{Derivative self-interactions for a massive vector field}},}\ }\href
  {\doibase 10.1016/j.physletb.2016.04.017} {\bibfield  {journal} {\bibinfo
  {journal} {Phys. Lett.}\ }\textbf {\bibinfo {volume} {B757}},\ \bibinfo
  {pages} {405--411} (\bibinfo {year} {2016})},\ \Eprint
  {http://arxiv.org/abs/1602.03410} {arXiv:1602.03410 [hep-th]} \BibitemShut
  {NoStop}%
%%CITATION = ARXIV:1602.03410;%%
\bibitem [{\citenamefont {De~Felice}\ \emph
  {et~al.}(2016{\natexlab{a}})\citenamefont {De~Felice}, \citenamefont
  {Heisenberg}, \citenamefont {Kase}, \citenamefont {Tsujikawa}, \citenamefont
  {Zhang},\ and\ \citenamefont {Zhao}}]{DeFelice:2016cri}%
  \BibitemOpen
  \bibfield  {author} {\bibinfo {author} {\bibfnamefont {Antonio}\ \bibnamefont
  {De~Felice}}, \bibinfo {author} {\bibfnamefont {Lavinia}\ \bibnamefont
  {Heisenberg}}, \bibinfo {author} {\bibfnamefont {Ryotaro}\ \bibnamefont
  {Kase}}, \bibinfo {author} {\bibfnamefont {Shinji}\ \bibnamefont
  {Tsujikawa}}, \bibinfo {author} {\bibfnamefont {Ying-li}\ \bibnamefont
  {Zhang}}, \ and\ \bibinfo {author} {\bibfnamefont {Gong-Bo}\ \bibnamefont
  {Zhao}},\ }\bibfield  {title} {\enquote {\bibinfo {title} {{Screening fifth
  forces in generalized Proca theories}},}\ }\href {\doibase
  10.1103/PhysRevD.93.104016} {\bibfield  {journal} {\bibinfo  {journal} {Phys.
  Rev.}\ }\textbf {\bibinfo {volume} {D93}},\ \bibinfo {pages} {104016}
  (\bibinfo {year} {2016}{\natexlab{a}})},\ \Eprint
  {http://arxiv.org/abs/1602.00371} {arXiv:1602.00371 [gr-qc]} \BibitemShut
  {NoStop}%
%%CITATION = ARXIV:1602.00371;%%
\bibitem [{\citenamefont {De~Felice}\ \emph
  {et~al.}(2016{\natexlab{b}})\citenamefont {De~Felice}, \citenamefont
  {Heisenberg}, \citenamefont {Kase}, \citenamefont {Mukohyama}, \citenamefont
  {Tsujikawa},\ and\ \citenamefont {Zhang}}]{DeFelice:2016yws}%
  \BibitemOpen
  \bibfield  {author} {\bibinfo {author} {\bibfnamefont {Antonio}\ \bibnamefont
  {De~Felice}}, \bibinfo {author} {\bibfnamefont {Lavinia}\ \bibnamefont
  {Heisenberg}}, \bibinfo {author} {\bibfnamefont {Ryotaro}\ \bibnamefont
  {Kase}}, \bibinfo {author} {\bibfnamefont {Shinji}\ \bibnamefont
  {Mukohyama}}, \bibinfo {author} {\bibfnamefont {Shinji}\ \bibnamefont
  {Tsujikawa}}, \ and\ \bibinfo {author} {\bibfnamefont {Ying-li}\ \bibnamefont
  {Zhang}},\ }\bibfield  {title} {\enquote {\bibinfo {title} {{Cosmology in
  generalized Proca theories}},}\ }\href {\doibase
  10.1088/1475-7516/2016/06/048} {\bibfield  {journal} {\bibinfo  {journal}
  {JCAP}\ }\textbf {\bibinfo {volume} {1606}},\ \bibinfo {pages} {048}
  (\bibinfo {year} {2016}{\natexlab{b}})},\ \Eprint
  {http://arxiv.org/abs/1603.05806} {arXiv:1603.05806 [gr-qc]} \BibitemShut
  {NoStop}%
%%CITATION = ARXIV:1603.05806;%%
\bibitem [{\citenamefont {De~Felice}\ \emph
  {et~al.}(2016{\natexlab{c}})\citenamefont {De~Felice}, \citenamefont
  {Heisenberg}, \citenamefont {Kase}, \citenamefont {Mukohyama}, \citenamefont
  {Tsujikawa},\ and\ \citenamefont {Zhang}}]{DeFelice:2016uil}%
  \BibitemOpen
  \bibfield  {author} {\bibinfo {author} {\bibfnamefont {Antonio}\ \bibnamefont
  {De~Felice}}, \bibinfo {author} {\bibfnamefont {Lavinia}\ \bibnamefont
  {Heisenberg}}, \bibinfo {author} {\bibfnamefont {Ryotaro}\ \bibnamefont
  {Kase}}, \bibinfo {author} {\bibfnamefont {Shinji}\ \bibnamefont
  {Mukohyama}}, \bibinfo {author} {\bibfnamefont {Shinji}\ \bibnamefont
  {Tsujikawa}}, \ and\ \bibinfo {author} {\bibfnamefont {Ying-li}\ \bibnamefont
  {Zhang}},\ }\bibfield  {title} {\enquote {\bibinfo {title} {{Effective
  gravitational couplings for cosmological perturbations in generalized Proca
  theories}},}\ }\href {\doibase 10.1103/PhysRevD.94.044024} {\bibfield
  {journal} {\bibinfo  {journal} {Phys. Rev.}\ }\textbf {\bibinfo {volume}
  {D94}},\ \bibinfo {pages} {044024} (\bibinfo {year} {2016}{\natexlab{c}})},\
  \Eprint {http://arxiv.org/abs/1605.05066} {arXiv:1605.05066 [gr-qc]}
  \BibitemShut {NoStop}%
%%CITATION = ARXIV:1605.05066;%%
\bibitem [{\citenamefont {Chagoya}\ \emph {et~al.}(2016)\citenamefont
  {Chagoya}, \citenamefont {Niz},\ and\ \citenamefont
  {Tasinato}}]{Chagoya:2016aar}%
  \BibitemOpen
  \bibfield  {author} {\bibinfo {author} {\bibfnamefont {Javier}\ \bibnamefont
  {Chagoya}}, \bibinfo {author} {\bibfnamefont {Gustavo}\ \bibnamefont {Niz}},
  \ and\ \bibinfo {author} {\bibfnamefont {Gianmassimo}\ \bibnamefont
  {Tasinato}},\ }\bibfield  {title} {\enquote {\bibinfo {title} {{Black Holes
  and Abelian Symmetry Breaking}},}\ }\href {\doibase
  10.1088/0264-9381/33/17/175007} {\bibfield  {journal} {\bibinfo  {journal}
  {Class. Quant. Grav.}\ }\textbf {\bibinfo {volume} {33}},\ \bibinfo {pages}
  {175007} (\bibinfo {year} {2016})},\ \Eprint
  {http://arxiv.org/abs/1602.08697} {arXiv:1602.08697 [hep-th]} \BibitemShut
  {NoStop}%
%%CITATION = ARXIV:1602.08697;%%
\bibitem [{\citenamefont {Minamitsuji}(2016)}]{Minamitsuji:2016ydr}%
  \BibitemOpen
  \bibfield  {author} {\bibinfo {author} {\bibfnamefont {Masato}\ \bibnamefont
  {Minamitsuji}},\ }\bibfield  {title} {\enquote {\bibinfo {title} {{Solutions
  in the generalized Proca theory with the nonminimal coupling to the Einstein
  tensor}},}\ }\href {\doibase 10.1103/PhysRevD.94.084039} {\bibfield
  {journal} {\bibinfo  {journal} {Phys. Rev.}\ }\textbf {\bibinfo {volume}
  {D94}},\ \bibinfo {pages} {084039} (\bibinfo {year} {2016})},\ \Eprint
  {http://arxiv.org/abs/1607.06278} {arXiv:1607.06278 [gr-qc]} \BibitemShut
  {NoStop}%
%%CITATION = ARXIV:1607.06278;%%
\bibitem [{\citenamefont {Geng}\ and\ \citenamefont {Lu}(2016)}]{Geng:2015kvs}%
  \BibitemOpen
  \bibfield  {author} {\bibinfo {author} {\bibfnamefont {Wei-Jian}\
  \bibnamefont {Geng}}\ and\ \bibinfo {author} {\bibfnamefont {H.}~\bibnamefont
  {Lu}},\ }\bibfield  {title} {\enquote {\bibinfo {title} {{Einstein-Vector
  Gravity, Emerging Gauge Symmetry and de Sitter Bounce}},}\ }\href {\doibase
  10.1103/PhysRevD.93.044035} {\bibfield  {journal} {\bibinfo  {journal} {Phys.
  Rev.}\ }\textbf {\bibinfo {volume} {D93}},\ \bibinfo {pages} {044035}
  (\bibinfo {year} {2016})},\ \Eprint {http://arxiv.org/abs/1511.03681}
  {arXiv:1511.03681 [hep-th]} \BibitemShut {NoStop}%
%%CITATION = ARXIV:1511.03681;%%
\bibitem [{\citenamefont {Chagoya}\ \emph {et~al.}(2017)\citenamefont
  {Chagoya}, \citenamefont {Niz},\ and\ \citenamefont
  {Tasinato}}]{Chagoya:2017fyl}%
  \BibitemOpen
  \bibfield  {author} {\bibinfo {author} {\bibfnamefont {Javier}\ \bibnamefont
  {Chagoya}}, \bibinfo {author} {\bibfnamefont {Gustavo}\ \bibnamefont {Niz}},
  \ and\ \bibinfo {author} {\bibfnamefont {Gianmassimo}\ \bibnamefont
  {Tasinato}},\ }\bibfield  {title} {\enquote {\bibinfo {title} {{Black Holes
  and Neutron Stars in Vector Galileons}},}\ }\href@noop {} {\  (\bibinfo
  {year} {2017})},\ \Eprint {http://arxiv.org/abs/1703.09555} {arXiv:1703.09555
  [gr-qc]} \BibitemShut {NoStop}%
%%CITATION = ARXIV:1703.09555;%%
\bibitem [{\citenamefont {Babichev}\ \emph {et~al.}(2017)\citenamefont
  {Babichev}, \citenamefont {Charmousis},\ and\ \citenamefont
  {Hassaine}}]{Babichev:2017rti}%
  \BibitemOpen
  \bibfield  {author} {\bibinfo {author} {\bibfnamefont {Eugeny}\ \bibnamefont
  {Babichev}}, \bibinfo {author} {\bibfnamefont {Christos}\ \bibnamefont
  {Charmousis}}, \ and\ \bibinfo {author} {\bibfnamefont {Mokhtar}\
  \bibnamefont {Hassaine}},\ }\bibfield  {title} {\enquote {\bibinfo {title}
  {{Black holes and solitons in an extended Proca theory}},}\ }\href {\doibase
  10.1007/JHEP05(2017)114} {\bibfield  {journal} {\bibinfo  {journal} {JHEP}\
  }\textbf {\bibinfo {volume} {05}},\ \bibinfo {pages} {114} (\bibinfo {year}
  {2017})},\ \Eprint {http://arxiv.org/abs/1703.07676} {arXiv:1703.07676
  [gr-qc]} \BibitemShut {NoStop}%
%%CITATION = ARXIV:1703.07676;%%
\bibitem [{\citenamefont {Heisenberg}\ \emph {et~al.}(2017)\citenamefont
  {Heisenberg}, \citenamefont {Kase}, \citenamefont {Minamitsuji},\ and\
  \citenamefont {Tsujikawa}}]{Heisenberg:2017xda}%
  \BibitemOpen
  \bibfield  {author} {\bibinfo {author} {\bibfnamefont {Lavinia}\ \bibnamefont
  {Heisenberg}}, \bibinfo {author} {\bibfnamefont {Ryotaro}\ \bibnamefont
  {Kase}}, \bibinfo {author} {\bibfnamefont {Masato}\ \bibnamefont
  {Minamitsuji}}, \ and\ \bibinfo {author} {\bibfnamefont {Shinji}\
  \bibnamefont {Tsujikawa}},\ }\bibfield  {title} {\enquote {\bibinfo {title}
  {{Hairy black-hole solutions in generalized Proca theories}},}\ }\href@noop
  {} {\  (\bibinfo {year} {2017})},\ \Eprint {http://arxiv.org/abs/1705.09662}
  {arXiv:1705.09662 [gr-qc]} \BibitemShut {NoStop}%
%%CITATION = ARXIV:1705.09662;%%
\bibitem [{\citenamefont {Kaup}(1968)}]{Kaup:1968zz}%
  \BibitemOpen
  \bibfield  {author} {\bibinfo {author} {\bibfnamefont {David~J.}\
  \bibnamefont {Kaup}},\ }\bibfield  {title} {\enquote {\bibinfo {title}
  {{Klein-Gordon Geon}},}\ }\href {\doibase 10.1103/PhysRev.172.1331}
  {\bibfield  {journal} {\bibinfo  {journal} {Phys. Rev.}\ }\textbf {\bibinfo
  {volume} {172}},\ \bibinfo {pages} {1331--1342} (\bibinfo {year}
  {1968})}\BibitemShut {NoStop}%
%%CITATION = PHRVA,172,1331;%%
\bibitem [{\citenamefont {Ruffini}\ and\ \citenamefont
  {Bonazzola}(1969)}]{Ruffini:1969qy}%
  \BibitemOpen
  \bibfield  {author} {\bibinfo {author} {\bibfnamefont {Remo}\ \bibnamefont
  {Ruffini}}\ and\ \bibinfo {author} {\bibfnamefont {Silvano}\ \bibnamefont
  {Bonazzola}},\ }\bibfield  {title} {\enquote {\bibinfo {title} {{Systems of
  selfgravitating particles in general relativity and the concept of an
  equation of state}},}\ }\href {\doibase 10.1103/PhysRev.187.1767} {\bibfield
  {journal} {\bibinfo  {journal} {Phys. Rev.}\ }\textbf {\bibinfo {volume}
  {187}},\ \bibinfo {pages} {1767--1783} (\bibinfo {year} {1969})}\BibitemShut
  {NoStop}%
%%CITATION = PHRVA,187,1767;%%
\bibitem [{\citenamefont {Friedberg}\ \emph {et~al.}(1987)\citenamefont
  {Friedberg}, \citenamefont {Lee},\ and\ \citenamefont
  {Pang}}]{Friedberg:1986tp}%
  \BibitemOpen
  \bibfield  {author} {\bibinfo {author} {\bibfnamefont {R.}~\bibnamefont
  {Friedberg}}, \bibinfo {author} {\bibfnamefont {T.~D.}\ \bibnamefont {Lee}},
  \ and\ \bibinfo {author} {\bibfnamefont {Y.}~\bibnamefont {Pang}},\
  }\bibfield  {title} {\enquote {\bibinfo {title} {{MINI - SOLITON STARS}},}\
  }\href {\doibase 10.1103/PhysRevD.35.3640} {\bibfield  {journal} {\bibinfo
  {journal} {Phys. Rev.}\ }\textbf {\bibinfo {volume} {D35}},\ \bibinfo {pages}
  {3640} (\bibinfo {year} {1987})},\ \bibinfo {note} {[,55(1986)]}\BibitemShut
  {NoStop}%
%%CITATION = PHRVA,D35,3640;%%
\bibitem [{\citenamefont {Jetzer}(1992)}]{Jetzer:1991jr}%
  \BibitemOpen
  \bibfield  {author} {\bibinfo {author} {\bibfnamefont {Philippe}\
  \bibnamefont {Jetzer}},\ }\bibfield  {title} {\enquote {\bibinfo {title}
  {{Boson stars}},}\ }\href {\doibase 10.1016/0370-1573(92)90123-H} {\bibfield
  {journal} {\bibinfo  {journal} {Phys. Rept.}\ }\textbf {\bibinfo {volume}
  {220}},\ \bibinfo {pages} {163--227} (\bibinfo {year} {1992})}\BibitemShut
  {NoStop}%
%%CITATION = PRPLC,220,163;%%
\bibitem [{\citenamefont {Schunck}\ and\ \citenamefont
  {Mielke}(2003)}]{Schunck:2003kk}%
  \BibitemOpen
  \bibfield  {author} {\bibinfo {author} {\bibfnamefont {F.E.}\ \bibnamefont
  {Schunck}}\ and\ \bibinfo {author} {\bibfnamefont {E.W.}\ \bibnamefont
  {Mielke}},\ }\bibfield  {title} {\enquote {\bibinfo {title} {{General
  relativistic boson stars}},}\ }\href {\doibase 10.1088/0264-9381/20/20/201}
  {\bibfield  {journal} {\bibinfo  {journal} {Class.Quant.Grav.}\ }\textbf
  {\bibinfo {volume} {20}},\ \bibinfo {pages} {R301--R356} (\bibinfo {year}
  {2003})},\ \Eprint {http://arxiv.org/abs/0801.0307} {arXiv:0801.0307
  [astro-ph]} \BibitemShut {NoStop}%
%%CITATION = ARXIV:0801.0307;%%
\bibitem [{\citenamefont {Gleiser}(1988)}]{Gleiser:1988rq}%
  \BibitemOpen
  \bibfield  {author} {\bibinfo {author} {\bibfnamefont {Marcelo}\ \bibnamefont
  {Gleiser}},\ }\bibfield  {title} {\enquote {\bibinfo {title} {{Stability of
  Boson Stars}},}\ }\href {\doibase 10.1103/PhysRevD.38.2376,
  10.1103/PhysRevD.39.1257} {\bibfield  {journal} {\bibinfo  {journal} {Phys.
  Rev.}\ }\textbf {\bibinfo {volume} {D38}},\ \bibinfo {pages} {2376} (\bibinfo
  {year} {1988})},\ \bibinfo {note} {[Erratum: Phys.
  Rev.D39,no.4,1257(1989)]}\BibitemShut {NoStop}%
%%CITATION = PHRVA,D38,2376;%%
\bibitem [{\citenamefont {Lee}\ and\ \citenamefont {Pang}(1989)}]{Lee:1988av}%
  \BibitemOpen
  \bibfield  {author} {\bibinfo {author} {\bibfnamefont {T.~D.}\ \bibnamefont
  {Lee}}\ and\ \bibinfo {author} {\bibfnamefont {Yang}\ \bibnamefont {Pang}},\
  }\bibfield  {title} {\enquote {\bibinfo {title} {{Stability of Mini - Boson
  Stars}},}\ }\href {\doibase 10.1016/0550-3213(89)90365-9} {\bibfield
  {journal} {\bibinfo  {journal} {Nucl. Phys.}\ }\textbf {\bibinfo {volume}
  {B315}},\ \bibinfo {pages} {477} (\bibinfo {year} {1989})},\ \bibinfo {note}
  {[,129(1988)]}\BibitemShut {NoStop}%
%%CITATION = NUPHA,B315,477;%%
\bibitem [{\citenamefont {Gleiser}\ and\ \citenamefont
  {Watkins}(1989)}]{Gleiser:1988ih}%
  \BibitemOpen
  \bibfield  {author} {\bibinfo {author} {\bibfnamefont {Marcelo}\ \bibnamefont
  {Gleiser}}\ and\ \bibinfo {author} {\bibfnamefont {Richard}\ \bibnamefont
  {Watkins}},\ }\bibfield  {title} {\enquote {\bibinfo {title} {{Gravitational
  Stability of Scalar Matter}},}\ }\href {\doibase
  10.1016/0550-3213(89)90627-5} {\bibfield  {journal} {\bibinfo  {journal}
  {Nucl. Phys.}\ }\textbf {\bibinfo {volume} {B319}},\ \bibinfo {pages}
  {733--746} (\bibinfo {year} {1989})}\BibitemShut {NoStop}%
%%CITATION = NUPHA,B319,733;%%
\bibitem [{\citenamefont {Hawley}\ and\ \citenamefont
  {Choptuik}(2000)}]{Hawley:2000dt}%
  \BibitemOpen
  \bibfield  {author} {\bibinfo {author} {\bibfnamefont {Scott~H.}\
  \bibnamefont {Hawley}}\ and\ \bibinfo {author} {\bibfnamefont {Matthew~W.}\
  \bibnamefont {Choptuik}},\ }\bibfield  {title} {\enquote {\bibinfo {title}
  {{Boson stars driven to the brink of black hole formation}},}\ }\href
  {\doibase 10.1103/PhysRevD.62.104024} {\bibfield  {journal} {\bibinfo
  {journal} {Phys. Rev.}\ }\textbf {\bibinfo {volume} {D62}},\ \bibinfo {pages}
  {104024} (\bibinfo {year} {2000})},\ \Eprint
  {http://arxiv.org/abs/gr-qc/0007039} {arXiv:gr-qc/0007039 [gr-qc]}
  \BibitemShut {NoStop}%
%%CITATION = GR-QC/0007039;%%
\bibitem [{\citenamefont {Brihaye}\ \emph {et~al.}(2016)\citenamefont
  {Brihaye}, \citenamefont {Cisterna},\ and\ \citenamefont
  {Erices}}]{Brihaye:2016lin}%
  \BibitemOpen
  \bibfield  {author} {\bibinfo {author} {\bibfnamefont {Yves}\ \bibnamefont
  {Brihaye}}, \bibinfo {author} {\bibfnamefont {Adolfo}\ \bibnamefont
  {Cisterna}}, \ and\ \bibinfo {author} {\bibfnamefont {Cristian}\ \bibnamefont
  {Erices}},\ }\bibfield  {title} {\enquote {\bibinfo {title} {{Boson stars in
  biscalar extensions of Horndeski gravity}},}\ }\href {\doibase
  10.1103/PhysRevD.93.124057} {\bibfield  {journal} {\bibinfo  {journal} {Phys.
  Rev.}\ }\textbf {\bibinfo {volume} {D93}},\ \bibinfo {pages} {124057}
  (\bibinfo {year} {2016})},\ \Eprint {http://arxiv.org/abs/1604.02121}
  {arXiv:1604.02121 [hep-th]} \BibitemShut {NoStop}%
%%CITATION = ARXIV:1604.02121;%%
\bibitem [{\citenamefont {Hartmann}\ \emph {et~al.}(2013)\citenamefont
  {Hartmann}, \citenamefont {Riedel},\ and\ \citenamefont
  {Suciu}}]{Hartmann:2013tca}%
  \BibitemOpen
  \bibfield  {author} {\bibinfo {author} {\bibfnamefont {Betti}\ \bibnamefont
  {Hartmann}}, \bibinfo {author} {\bibfnamefont {Jurgen}\ \bibnamefont
  {Riedel}}, \ and\ \bibinfo {author} {\bibfnamefont {Raluca}\ \bibnamefont
  {Suciu}},\ }\bibfield  {title} {\enquote {\bibinfo {title} {{Gauss-Bonnet
  boson stars}},}\ }\href {\doibase 10.1016/j.physletb.2013.09.050} {\bibfield
  {journal} {\bibinfo  {journal} {Phys. Lett.}\ }\textbf {\bibinfo {volume}
  {B726}},\ \bibinfo {pages} {906--912} (\bibinfo {year} {2013})},\ \Eprint
  {http://arxiv.org/abs/1308.3391} {arXiv:1308.3391 [gr-qc]} \BibitemShut
  {NoStop}%
%%CITATION = ARXIV:1308.3391;%%
\bibitem [{\citenamefont {Brihaye}\ and\ \citenamefont
  {Riedel}(2014)}]{Brihaye:2013zha}%
  \BibitemOpen
  \bibfield  {author} {\bibinfo {author} {\bibfnamefont {Yves}\ \bibnamefont
  {Brihaye}}\ and\ \bibinfo {author} {\bibfnamefont {Jurgen}\ \bibnamefont
  {Riedel}},\ }\bibfield  {title} {\enquote {\bibinfo {title} {{Rotating boson
  stars in five-dimensional Einstein-Gauss-Bonnet gravity}},}\ }\href {\doibase
  10.1103/PhysRevD.89.104060} {\bibfield  {journal} {\bibinfo  {journal} {Phys.
  Rev.}\ }\textbf {\bibinfo {volume} {D89}},\ \bibinfo {pages} {104060}
  (\bibinfo {year} {2014})},\ \Eprint {http://arxiv.org/abs/1310.7223}
  {arXiv:1310.7223 [gr-qc]} \BibitemShut {NoStop}%
%%CITATION = ARXIV:1310.7223;%%
\bibitem [{\citenamefont {Baibhav}\ and\ \citenamefont
  {Maity}(2017)}]{Baibhav:2016fot}%
  \BibitemOpen
  \bibfield  {author} {\bibinfo {author} {\bibfnamefont {Vishal}\ \bibnamefont
  {Baibhav}}\ and\ \bibinfo {author} {\bibfnamefont {Debaprasad}\ \bibnamefont
  {Maity}},\ }\bibfield  {title} {\enquote {\bibinfo {title} {{Boson Stars in
  Higher Derivative Gravity}},}\ }\href {\doibase 10.1103/PhysRevD.95.024027}
  {\bibfield  {journal} {\bibinfo  {journal} {Phys. Rev.}\ }\textbf {\bibinfo
  {volume} {D95}},\ \bibinfo {pages} {024027} (\bibinfo {year} {2017})},\
  \Eprint {http://arxiv.org/abs/1609.07225} {arXiv:1609.07225 [gr-qc]}
  \BibitemShut {NoStop}%
%%CITATION = ARXIV:1609.07225;%%
\bibitem [{\citenamefont {Brito}\ \emph {et~al.}(2016)\citenamefont {Brito},
  \citenamefont {Cardoso}, \citenamefont {Herdeiro},\ and\ \citenamefont
  {Radu}}]{Brito:2015pxa}%
  \BibitemOpen
  \bibfield  {author} {\bibinfo {author} {\bibfnamefont {Richard}\ \bibnamefont
  {Brito}}, \bibinfo {author} {\bibfnamefont {Vitor}\ \bibnamefont {Cardoso}},
  \bibinfo {author} {\bibfnamefont {Carlos A.~R.}\ \bibnamefont {Herdeiro}}, \
  and\ \bibinfo {author} {\bibfnamefont {Eugen}\ \bibnamefont {Radu}},\
  }\bibfield  {title} {\enquote {\bibinfo {title} {{Proca stars: Gravitating
  Bose-Einstein condensates of massive spin 1 particles}},}\ }\href {\doibase
  10.1016/j.physletb.2015.11.051} {\bibfield  {journal} {\bibinfo  {journal}
  {Phys. Lett.}\ }\textbf {\bibinfo {volume} {B752}},\ \bibinfo {pages}
  {291--295} (\bibinfo {year} {2016})},\ \Eprint
  {http://arxiv.org/abs/1508.05395} {arXiv:1508.05395 [gr-qc]} \BibitemShut
  {NoStop}%
%%CITATION = ARXIV:1508.05395;%%
\bibitem [{\citenamefont {Landea}\ and\ \citenamefont
  {Garcia}(2016)}]{Garcia:2016ldc}%
  \BibitemOpen
  \bibfield  {author} {\bibinfo {author} {\bibfnamefont {Ignacio~Salazar}\
  \bibnamefont {Landea}}\ and\ \bibinfo {author} {\bibfnamefont {Federico}\
  \bibnamefont {Garcia}},\ }\bibfield  {title} {\enquote {\bibinfo {title}
  {{Charged Proca Stars}},}\ }\href {\doibase 10.1103/PhysRevD.94.104006}
  {\bibfield  {journal} {\bibinfo  {journal} {Phys. Rev.}\ }\textbf {\bibinfo
  {volume} {D94}},\ \bibinfo {pages} {104006} (\bibinfo {year} {2016})},\
  \Eprint {http://arxiv.org/abs/1608.00011} {arXiv:1608.00011 [hep-th]}
  \BibitemShut {NoStop}%
%%CITATION = ARXIV:1608.00011;%%
\bibitem [{\citenamefont {Brihaye}\ \emph {et~al.}(2017)\citenamefont
  {Brihaye}, \citenamefont {Delplace},\ and\ \citenamefont
  {Verbin}}]{Brihaye:2017inn}%
  \BibitemOpen
  \bibfield  {author} {\bibinfo {author} {\bibfnamefont {Y.}~\bibnamefont
  {Brihaye}}, \bibinfo {author} {\bibfnamefont {Th.}\ \bibnamefont {Delplace}},
  \ and\ \bibinfo {author} {\bibfnamefont {Y.}~\bibnamefont {Verbin}},\
  }\bibfield  {title} {\enquote {\bibinfo {title} {{Proca Q Balls and their
  Coupling to Gravity}},}\ }\href@noop {} {\  (\bibinfo {year} {2017})},\
  \Eprint {http://arxiv.org/abs/1704.01648} {arXiv:1704.01648 [gr-qc]}
  \BibitemShut {NoStop}%
%%CITATION = ARXIV:1704.01648;%%
\bibitem [{\citenamefont {Woodard}(2015)}]{Woodard:2015zca}%
  \BibitemOpen
  \bibfield  {author} {\bibinfo {author} {\bibfnamefont {Richard~P.}\
  \bibnamefont {Woodard}},\ }\bibfield  {title} {\enquote {\bibinfo {title}
  {{Ostrogradsky's theorem on Hamiltonian instability}},}\ }\href {\doibase
  10.4249/scholarpedia.32243} {\bibfield  {journal} {\bibinfo  {journal}
  {Scholarpedia}\ }\textbf {\bibinfo {volume} {10}},\ \bibinfo {pages} {32243}
  (\bibinfo {year} {2015})},\ \Eprint {http://arxiv.org/abs/1506.02210}
  {arXiv:1506.02210 [hep-th]} \BibitemShut {NoStop}%
%%CITATION = ARXIV:1506.02210;%%
\bibitem [{\citenamefont {Derrick}(1964)}]{Derrick:1964ww}%
  \BibitemOpen
  \bibfield  {author} {\bibinfo {author} {\bibfnamefont {G.~H.}\ \bibnamefont
  {Derrick}},\ }\bibfield  {title} {\enquote {\bibinfo {title} {{Comments on
  nonlinear wave equations as models for elementary particles}},}\ }\href
  {\doibase 10.1063/1.1704233} {\bibfield  {journal} {\bibinfo  {journal} {J.
  Math. Phys.}\ }\textbf {\bibinfo {volume} {5}},\ \bibinfo {pages}
  {1252--1254} (\bibinfo {year} {1964})}\BibitemShut {NoStop}%
%%CITATION = JMAPA,5,1252;%%
\end{thebibliography}%
\end{document}